%% file: arXiv3.tex
\documentclass[a4paper,notitlepage]{report}   	%
\renewcommand{\thechapter}{{\textbf{\arabic{chapter}}}}
\renewcommand{\thesection}{\thechapter:$\,$\arabic{section}}

\renewcommand{\appendix}{
	\section*{Appendices}
	\setcounter{section}{0}
	\renewcommand{\thesection}{\thechapter:$\,$\Alph{section}}
}
\newcommand{\appendixend}{
	\setcounter{section}{0}
	\renewcommand{\thesection}{\thechapter:$\,$\arabic{section}}
}
\usepackage{amsmath}
\usepackage{amssymb}
\usepackage{xr}
\input{Parts/Commands.tex}
\newcommand{\chap}[1]{}
\newcommand{\notchap}[1]{#1}

\newcommand{\notstandalone}[1]{}
\usepackage{geometry}
\usepackage[numbers,square,sort&compress]{natbib}

\begin{document}

\title{A Classical Analogue to the Standard Model\\and General Relativity}
\author{Chapter 3\\~\\R. N. C. Pfeifer}

\date{01 January 2024}

\maketitle
\thispagestyle{empty}
\newpage

\tableofcontents

\setcounter{chapter}{2}
\input{Parts/Chapter3.tex}

\bibliographystyle{apsrev4-2}
\bibliography{Paper.bib}

\end{document}

%% file: Parts/Commands.tex
\usepackage{refcount}

\usepackage{graphicx}
\usepackage{mathrsfs}
\usepackage{bm}
\usepackage[hyperref]{xcolor} %
\definecolor{darkgreen}{rgb}{0,0.4,0} %
\definecolor{darkred}{rgb}{0.55,0,0} %
\definecolor{navy}{rgb}{0,0,0.55} %
\usepackage[hypertexnames=false,breaklinks=true,colorlinks=true,citecolor=darkgreen,linkcolor=black,urlcolor=navy]{hyperref} %
\usepackage{amsthm}
\usepackage{lscape}
\theoremstyle{definition} %

\newcommand{\G}{\mc{G}}
\newcommand{\Z}{\mc{Z}}

\newcommand{\prm}[1]{\protect{$#1$}}

\newcommand{\sol}[1]{}
\newcommand{\h}{\mathrm{H}}%
\newcommand{\bmh}{{\bm{\mathrm{H}}}}%

\newcommand{\ta}{{\tilde{a}}}
\newcommand{\tb}{{\tilde{b}}}
\newcommand{\tc}{{\tilde{c}}}
\newcommand{\td}{{\tilde{d}}}

\newcommand{\bma}{{\bm{a}}}
\newcommand{\bmb}{{\bm{b}}}
\newcommand{\bmc}{{\bm{c}}}

\newcommand{\tta}{{\tilde{a}'}}
\newcommand{\ttb}{{\tilde{b}'}}
\newcommand{\ttc}{{\tilde{c}'}}
\newcommand{\ttd}{{\tilde{d}'}}
\newcommand{\ulambda}{\lambda} %
\newcommand{\mrm}[1]{\mathrm{#1}}
\newcommand{\mbf}[1]{\mathbf{#1}}
\newcommand{\mbb}[1]{\mathbb{#1}}
\newcommand{\mc}[1]{\mathcal{#1}}
\newcommand{\mfk}[1]{\mathfrak{#1}}
\newcommand{\mscr}[1]{\mathscr{#1}}
\newcommand{\bgrid}{\left(\begin{array}{rrr}}
\newcommand{\egrid}{\end{array}\right)}
\newcommand{\bgridt}{\left(\begin{array}{rr}}
\newcommand{\egridt}{\end{array}\right)}
\newcommand{\bgridtt}{\left[\begin{array}{rr}}
\newcommand{\egridtt}{\end{array}\right]}
\newcommand{\os}{\!\!\not\!} %
\newcommand{\eref}[1]{(\ref{#1})}
\newcommand{\Eref}[1]{Eq.~(\ref{#1})}

\newcommand{\Erefr}[2]{Eqs.~(\ref{#1}--\ref{#2})}
\newcommand{\Eqrefr}[2]{Equations~(\ref{#1})--(\ref{#2})}
\newcommand{\erefr}[2]{(\ref{#1}--\ref{#2})}

\newcommand{\Erefs}[2]{Eqs.~(\ref{#1}) and~(\ref{#2})}

\newcommand{\erefs}[2]{(\ref{#1}) and~(\ref{#2})}
\newcommand{\ereft}[2]{(\ref{#1},\ref{#2})}
\newcommand{\cref}[1]{Chapter~\ref{#1}}
\newcommand{\Cref}[1]{Chapter~\ref{#1}}
\newcommand{\crefs}[2]{Chapters~\ref{#1} and~\ref{#2}}
\newcommand{\crefr}[2]{Chapters~\ref{#1}--\ref{#2}}

\newcommand{\Peref}[2]{(\ref{#2})}
\newcommand{\PEref}[2]{Eq.~(\ref{#2})}

\newcommand{\Perefr}[3]{(\ref{#2}--\ref{#3})}

\newcommand{\PEreft}[2]{Eqs.~(\ref{#2})}

\newcommand{\fref}[1]{Fig.~\ref{#1}}

\newcommand{\freft}[1]{Figs.~\ref{#1}}
\newcommand{\sref}[1]{Sec.~\ref{#1}}
\newcommand{\Psref}[2]{\sref{#2}}

\newcommand{\srefs}[2]{Secs.~\ref{#1} and~\ref{#2}}
\newcommand{\sreft}[1]{Secs.~\ref{#1}}

\newcommand{\srefr}[2]{Secs.~\ref{#1}--\ref{#2}}
\newcommand{\aref}[1]{Appendix~\ref{#1}}
\newcommand{\arefr}[2]{Appendices~\ref{#1}--\ref{#2}}
\newcommand{\tref}[1]{Table~\ref{#1}}
\newcommand{\bsm}{\bar\sigma^\mu}
\newcommand{\bsmm}{\bar\sigma_\mu}
\newcommand{\bsn}{\bar\sigma^\nu}

\newcommand{\nn}{\nonumber}

\newcommand{\rmi}{\mathrm{i}}%
\newcommand{\la}{\langle}
\newcommand{\ra}{\rangle}
\newcommand{\dslash}{{\,\os\partial}}
\newcommand{\Dslash}{{\,\os\! D}}

\newcommand{\bdag}{{(\dagger)}}
\newcommand{\Tr}{\mrm{Tr}\,}

\newcommand{\rcite}[1]{Ref.~\cite{#1}}

\newcommand{\p}[1]{\phantom{#1}}

\newcommand{\Weyll}[2]{\left[\begin{aligned}&#1\\&#2\end{aligned}\right]}
\newcommand{\triplet}[3]{\left(\begin{aligned}&#1\\&#2\\&#3\end{aligned}\right)}

\newcommand{\fg}{\mrm{fg}}
\newcommand{\bgfield}[1]{[#1]_\mrm{bg}}
\newcommand{\fgfield}[1]{[#1]_\fg}
\newcommand{\OO}[1]{\mrm{O}\left(#1\right)}

\newcommand{\SU}[1]{\mrm{SU}(#1)}
\newcommand{\su}[1]{\mrm{su}(#1)}
\newcommand{\SO}[1]{\mrm{SO}(#1)}

\newcommand{\U}[1]{\mrm{U}(#1)}

\newcommand{\rmU}{\mrm{U}}

\newcommand{\peref}[1]{\protect{(\ref{#1})}}
\newcommand{\pEref}[1]{\protect{Eq.~(\ref{#1})}}

\newcommand{\psref}[1]{\protect{Sec.~\ref{#1}}}

\newcommand{\GL}[2]{\mrm{GL}(#1,\mbb{#2})}

\newcommand{\SL}[2]{\mrm{SL}(#1,\mbb{#2})}

\newcommand{\GLNR}{\GL{9}{R}}%

\renewcommand{\bar}[1]{\overline{#1}}

\newcommand{\preon}{\psi}
\newcommand{\quark}{q}

\newcommand{\ILO}[1]{\mrm{O}(#1)}

\newcommand{\bmmf}{\tilde{f}}

\newcommand{\N}{{\mathfrak{n}}}%
\newcommand{\RM}{{{\mbb{R}^{1,3}}}}

\newcommand{\Cw}[1]{{{\mbb{C}^{\wedge #1}}}}

\newcommand{\vp}{\varphi}
\newcommand{\tvp}{{\tilde{\varphi}}}
\newcommand{\bmvp}{{\bm{\vp}}}

\newcommand{\Cwt}{{\Cw{2}}}

\newcommand{\Cwtn}{{\Cw{2\N}}}

\newcommand{\Ptagref}[2]{{\tag{\ref{#2}}}}%

\newcommand{\quietGversion}[1]{}
\newcommand{\onlyinsummary}[1]{}

\newcommand{\schapnotchap}[2]{(\notchap{\rcite{#2},~}\sref{#1})}

\newcommand{\completed}[1]{}
\newcommand{\chapeight}[1]{}

%% file: Parts/Chapter3.tex
\chapter{Standard Model particle spectrum from scalar fields on \protect{$\mathbb{C}^{\wedge 18}$}\label{ch:SM}}

\begin{abstract}
The $\Cwtn$ models are analogue models which generate Lagrangians for quasiparticles on $\RM$ from antisymmetric vector products on Grassmann manifolds. This chapter introduces $\Cw{18}$, the smallest member of this series which is capable of hosting a quasiparticle spectrum analogous to the Standard Model. Once all gaugeable degrees of freedom have been fixed, the particle spectrum of $\Cw{18}$ is seen to resemble the Standard Model plus two additional weakly interacting bosons and a ninth gluon.
\end{abstract}

\section{Introduction}

The idea behind an analogue model \cite{maynard2001,dragoman2004,lewenstein2007} is one of the most fundamental concepts in modern physics---that different systems whose behaviour is governed by the same equations will exhibit equivalent behaviours, within the range of validity of the correspondence. Analogue models may be either thought experiments or laboratory realisations, and range from the simple to the sublime. They may be as basic as mechanics without friction, as specific as solving the one-dimensional quantum Ising chain by mapping to the two-dimensional classical Ising model \cite{onsager1944,suzuki1976}, or as broad as the study of gravitational analogues in an incredibly wide range of substrates \cite{visser2002,liberati2009,barcelo2011,unruh1981,garay2000,garay2001,lahav2010,gordon1923,leonhardt1999,leonhardt2000,jacobson1998,reznik2000,schutzhold2005,schutzhold2002}. 

The $\Cwtn$ series of models, introduced in \crefr{ch:simplest}{ch:colour}, are classical models which emulate quantum field theories (QFTs) on $\RM$ in their low-energy regimes. The complexity of the theory emulated, and the number of species it contains, increases with the increasing value of $\N$. These models may be considered analogue models to the QFTs which they emulate. %

For small $\N$, analogies between the $\Cwtn$ models and simple, easily-understood quantum systems permit insight into the behaviours of the quasiparticle excitations of the $\Cwtn$ models. For example, in \cref{ch:colour} the emergence of a local $\SU{3}$ symmetry %
implied binding of preon triplets to yield composite fermions in the low-energy regime. 
But for larger $\N$, the benefit of these models comes when they encompass QFTs which are themselves incompletely understood.
It is then interesting to probe the behaviours of these analogue systems %
and assess to what extent these might represent real phenomena in the emulated QFT.

The smallest member of the $\Cwtn$ series which is capable of encompassing the Standard Model is $\Cw{18}$. The present chapter introduces this model, identifies and fixes its gauge symmetries, and shows that the resulting particle spectrum resembles that of the Standard Model plus three additional vector bosons. (Two of these are later eliminated in \cref{ch:gravity}, and the third is a dark matter candidate.) In regimes in which these extra bosons may be ignored, model $\Cw{18}$ may therefore be viewed as an analogue of the Standard Model. In determining the faithfulness of such an analogue model, key questions include the strength and character of its interactions and the masses of the represented species. This chapter makes a start on exploring the analogy, constructing the interaction Lagrangian with zeroth-order evaluation of coupling coefficients. This is followed in subsequent chapters by zeroth-order evaluations of the boson and fermion rest masses (\crefs{ch:fermion}{ch:boson}), and then by their first-order refinements (\cref{ch:detail}). %

\section{Conventions}

This chapter follows the same conventions as \crefr{ch:simplest}{ch:colour}. Units are chosen such that $c=1,~h=1$.
On spinors, $S$-type indices are labelled from the beginning of the greek alphabet while $R$-type indices are labelled from the roman alphabet.
When equations and lemmas from \crefs{ch:simplest}{ch:colour} are referenced, they take the forms (\textbf{1}.1), (\textbf{2}.1), %
etc.

For brevity, given a complex boson $b_\mu$, the notation $b^\bdag_\mu$ is to be read as ``$b_\mu$ or $b^\dagger_\mu$''.

Regarding terminology around Feynman diagrams and symmetry factors:
\begin{itemize}
\item Where there exist multiple ways to connect up sources, vertices, and sinks to obtain equivalent diagrams up to interchange of %
co-ordinates on equivalent fields, the same term is obtained from the generator $\Z$ in multiple different ways and thus the diagram acquires a multiplicative factor. This is referred to in the present volume as a \emph{symmetry factor}.
\item Where integration over the parameters of a diagram (for example, over source/sink co-ordinates) 
yields the same diagram multiple times up to interchange of labels on these parameters, 
this represents a double- (or multiple-)counting of physical processes. It is then necessary to eliminate this multiple-counting by dividing by the appropriate %
factor. This is referred to in the present volume as \emph{diagrammatic redundancy} or \emph{double- \mbox{(multiple-)counting.}} In other sources these factors are also sometimes called symmetry factors, but in the present monograph for clarity this terminology is reserved for factors greater than one.
\end{itemize}

\section{The model}

\subsection{Symmetries of \protect{$\Cw{18}$}\label{sec:symC18}}

As in \cref{ch:colour}, the natural place to start the exploration of manifold $\Cw{18}$ is with its symmetries. Following \sref{sec:C2nsym}, the global symmetry of $\Cw{18}$ is
\begin{equation}
\GL{18}{C}\oplus T^{18}_{\Cw{18}}
\end{equation}
of which $\GL{18}{C}$ admits the decomposition
\begin{align}
\GL{18}{C}&\cong\GL{9}{R}\otimes\GL{2}{C}\label{eq:GL18Cdecomp1}\\
&\cong\left[\SU{9}\oplus\GL{1}{R}\right]\otimes\left[\SL{2}{C}\oplus\mbb{C}^\times\right].\nn
\end{align}
However, this may be further rewritten by applying Lemma~\Peref{II}{Lem:12} to obtain
\begin{equation}
\GL{18}{C}\cong\GL{3}{R}_A\otimes\GL{3}{R}_C\otimes\GL{2}{C},\label{eq:GL18Cdecomp2pre}
\end{equation}
where labels $A$ and $C$ distinguish the two copies of $\GL{3}{R}$.
Applying Lemmas~\Peref{II}{Lem:1} and~\Peref{II}{Lem:9} consecutively to subgroup $\GL{3}{R}_A\otimes\GL{2}{C}$ and then to $\GL{3}{R}_C\otimes\GL{2}{C}$, followed by Lemma~\Peref{II}{Lem:4}, then yields
\begin{equation}
\begin{split}
\GL{18}{C}\cong&\p{\otimes}[\SU{3}_A\oplus\GL{1}{R}_A]\\&\otimes[\SU{3}_C\oplus\GL{1}{R}_C]\\&\otimes[\SL{2}{C}\oplus\mbb{C}^\times].
\end{split}\label{eq:GL18Cdecomp2}
\end{equation}

As previously, selecting a submanifold $\Cwt\subset\Cw{18}$, denoted $\Cw{2\bullet}$, permits introduction of a 1:1 mapping~$\G$ from a subspace $M\subset\Cw{2\bullet}$ to $\RM$, and the translation group $T^{18}_\Cw{18}$ contains a subgroup $T^{2}_M$ which maps to $T^4_\RM$ under the action of $\G$. Once again a maximum-entropy pseudovacuum is introduced, comprising an infinite number of finitely spaced unitless scalar fields on $\Cw{18}$, and a product field is constructed on $\RM$. 
Starting from \Eref{eq:GL18Cdecomp1} and following \cref{ch:colour} then yields field content on $\RM$ comprising nine preons $\psi^{n\alpha}$ bearing an $R$-index $n\in\{1,\ldots,9\}$, eighty-one vector bosons $\vp^{\dot nn}_\mu$ which promote the global $\GL{9}{R}$ symmetry of \Eref{eq:GL18Cdecomp1} to a local symmetry on the foreground fields, and a complex scalar boson $\h$. Construction of an effective derivative operator for foreground fields on $\RM$ yields
\begin{align}
{D}_\mu=\partial_\mu&-\rmi fe_{\dot nn}\fgfield{\vp^{\dot nn}_\mu- \rmi{f}\bar\psi^{\dot n}\bsmm\psi^{n}}\label{eq:expandedD81}\\
&-\frac{\rmi f}{2}\fgfield{\Upsilon_\mu(\h+\rmi f\psi^a\psi_a)+\bar\Upsilon_\mu(\h^*+\rmi f\bar\psi_{\dot a}\bar\psi^{\dot a})}.\nn
\end{align}
Anticipating the persistence of an unbroken emergent local $\SU{3}$ symmetry, confinement of preons permits this to be reduced to
\begin{align}
{D}_\mu=\partial_\mu&-\rmi \bmmf e_{\dot nn}\fgfield{\bmvp^{\dot nn}_\mu}-\frac{\rmi \bmmf'}{2}\fgfield{\Upsilon_\mu\bmh+\bar\Upsilon_\mu\bmh^*}\label{eq:reducedD81}
\end{align}
at energies small compared to the preon confinement scale $\mc{E}_\preon$, with values of $\bmmf$ and $\bmmf'$ determined in \srefr{sec:photonint}{sec:weakint} and~\ref{sec:scalbosint} respectively. As in \cref{ch:colour}, bold fields correspond to preon pairs bound by an $\SU{3}$ local symmetry to be elucidated below. The operator $D_\mu$ may be understood as a covariant derivative with respect to $\GL{9}{R}$-valued cotangent and $\mbb{C}^\times$ bundles over $\RM$, with the representation of $\GL{9}{R}$ over $\mbb{C}^\times$ being trivial.

Decomposition~\eref{eq:GL18Cdecomp2} provides a re-expression of $\GL{9}{R}$ as the product of two dimension-9 subgroups, %
\begin{align}
\begin{split}
\GL{9}{R}=\;&[\SU{3}_A\oplus\GL{1}{R}_A]\\&\otimes [\SU{3}_C\oplus\GL{1}{R}_C]
\end{split}\label{eq:GL9Rdecomp}\\
\bm{81}=\;&[\bm{8}+\bm{1}]\times[\bm{8}+\bm{1}],
\end{align}
and---as seen in \aref{apdx:gpfac}%
---in the continuum limit the eighty-one vector bosons $\bmvp^{\dot nn}_\mu$ %
may then be replaced by two sets of nine vector bosons, one associated with $[\SU{3}_A\oplus\GL{1}{R}_A]$ and one with $[\SU{3}_C\oplus\GL{1}{R}_C]$.
The charges on the preons then %
admit the %
decomposition
\begin{equation}
n\rightarrow(a,c)\qquad\begin{array}{rl}
n&\in\{1,\ldots,9\}\\
a&\in\{1,2,3\}\\
c&\in\{1,2,3\}~\mrm{equivalently}~\{r,g,b\}
\end{array}\label{eq:glnrIndexDecomp}
\end{equation}
and the action of an arbitrary boson field $b^\tb_\mu\in\GLNR$ may be replaced by the joint action of a pair of fields $a^\ta_\mu\in[\SU{3}_A\otimes\GL{1}{R}_C]$ and $c^\tc_\mu\in[\SU{3}_C\oplus\GL{1}{R}_C]$ where the values of $\tb$ enumerate the pairs $(\ta,\tc)$.
It is worth noting that this reduction in boson number breaks down for a quantised model in the few-particle regime, as it replaces a single boson $b^\tb_\mu$ with a pair of bosons $a^\ta_\mu$ and $c^\tc_\mu$. Since the $\Cwtn$ model series emulates a quantised model, additional considerations are therefore required in the small particle number regime. These are discussed in \aref{apdx:gaugeSU9} and \crefs{ch:detail}{ch:CDF2} (in particular \aref{apdx:ACseparability}). %
Fortunately, the consequences of these considerations are limited to relatively subtle effects involving the weak vector bosons, and may for the moment be safely neglected.

Decomposing the $\GLNR$ index as per \Eref{eq:glnrIndexDecomp}, the associated preon and vector boson fields may be written
\begin{equation}
\begin{split}
\psi^{ac}_{\alpha}\qquad &a\in\{1,2,3\}\quad c\in\{r,g,b\}\\
\fgfield{\bma^\ta_\mu}\qquad &\ta\in\{1,\ldots,9\}\\
\fgfield{\bmc^\tc_\mu}\qquad &\tc\in\{1,\ldots,9\}
\end{split}
\end{equation}
where $\fgfield{\bma^\ta_\mu}$ and $\fgfield{\bmc^\tc_\mu}$ are composite bosons in the style of $\bmvp^{\dot nn}_\mu$, being made up of a pair of preons which may be separated by $\ILO{\mc{L}_\preon}$ in the pseudovacuum isotropy frame, with the index of the unexposed sector being contracted in an Einstein sum. Also introduce the background fields $\bgfield{a^\ta_\mu}$ and $\bgfield{c^\tc_\mu}$ having the same structure as their foreground counterparts, but with all partial derivatives required to act at the same point on the same fundamental scalar field (FSF) $\vp_q(x)$ as per \sref{sec:bgfieldsequiv}:
\begin{align}
&\bgfield{a^\ta_\mu}\lambda^A_\ta:=\delta_{\dot cc}\bar\partial^{A\dot c}\bsmm\lambda^A_\ta\partial^{Ac}\,\vp(x)\\
&\bgfield{c^\tc_\mu}\lambda^C_\tc:=\delta_{\dot aa}\bar\partial^{C\dot a}\bsmm\lambda^C_\tc\partial^{Ca}\,\vp(x)\\
&~\partial^A=\triplet{\partial^{1c}}{\partial^{2c}}{\partial^{3c}}\qquad \partial^C=\triplet{\partial^{ar}}{\partial^{ag}}{\partial^{ab}}.
\end{align}
The emergent covariant
derivative for the foreground fields then takes the form %
\begin{align}
\begin{split}
{D}_\mu=\partial_\mu&-\rmi \bmmf\ulambda^A_\ta\fgfield{\bma^{\ta}_\mu}-\rmi \bmmf\ulambda^C_\tc\fgfield{\bmc^{\tc}_\mu}\\
&-\frac{\rmi \bmmf'}{2}\fgfield{\Upsilon_\mu\bmh+\bar\Upsilon_\mu\bmh^*},
\end{split}
\label{eq:III:expandedD}
\end{align}
where $\ulambda^B_\tb$, $B\in\{A,C\}$, $\tb\in\{1,\ldots,8\}$ are the rescaled Gell-Mann matrices \Perefr{II}{eq:Cbasis1}{eq:Cbasis2} satisfying 
\begin{equation}
\Tr{\{\ulambda_\ta^B\ulambda^B_\tb\}}=\delta_{\ta\tb}
\end{equation}
and $\ulambda^B_9$ satisfies
\begin{equation}
[\ulambda^B_9]_{\dot bb} = \frac{1}{\sqrt{3}}\delta_{\dot bb}.\label{eq:lambda9}
\end{equation}
This is the covariant derivative of a local $[\SU{3}_A\oplus\GL{1}{R}_A]\otimes[\SU{3}_C\oplus\GL{1}{R}_C]$ symmetry over the tangent and complex scalar bundles, with both $[\SU{3}_B\oplus\GL{1}{R}_B]$, $B\in\{A,C\}$ having trivial representation over the complex scalar bundle.

Next, observe that the matrix representations %
$\lambda^A_9$ and $\lambda^C_9$ are identical in their actions on both the $A$ and the $C$ sectors, both corresponding to
\begin{equation}
\frac{1}{\sqrt{3}}\left(\mbb{I}_A\otimes\mbb{I}_C\right)=:\ulambda^N.
\end{equation}
This accidental degeneracy permits the replacement
\begin{equation}
\begin{split}
&\fgfield{\bma^9_\mu}\ulambda^A_9+\fgfield{\bmc^9_\mu}\ulambda^C_9\longrightarrow \fgfield{N_\mu} \ulambda^N\\
&\bgfield{a^9_\mu}\ulambda^A_9+\bgfield{c^9_\mu}\ulambda^C_9\longrightarrow \bgfield{N_\mu} \ulambda^N,
\end{split}
\end{equation}
defining a vector boson $N_\mu$ and reducing the local symmetry to $\SU{3}_A\oplus\SU{3}_C\oplus\GL{1}{R}_N$ explicitly over the tangent bundle and implicitly over the complex scalar bundle.\footnote{While spatially extensive composite bosons %
are generally denoted in bold (e.g.~\prm{\fgfield{\bmvp_\mu}}) to distinguish them from their pointlike counterparts (e.g.~\prm{\bgfield{\vp_\mu}}) and high-energy regime counterparts (e.g.~\prm{\tvp_\mu}), an exception is made for named bosons of the low-energy limit, e.g.~\prm{\fgfield{W_\mu}}, as the foreground field is always spatially extensive and the background field is not. Their spatially extensive forms are therefore left unbolded to simplify notation.} 
The covariant derivative below the preon scale therefore becomes 
\begin{align}
{D}_\mu=\partial_\mu&-\rmi \bmmf\ulambda^A_\tta\fgfield{\bma^{\tta}_\mu}-\rmi \bmmf\ulambda^C_{\ttc}\fgfield{\bmc^{\ttc}_\mu}-\rmi \bmmf \ulambda^N\fgfield{N_\mu}\nn\\
&-\frac{\rmi \bmmf'}{2}\fgfield{\Upsilon_\mu\bmh+\bar\Upsilon_\mu\bmh^*}
\label{eq:reducedDwithN} %
\end{align}
where $\tta$ and $\ttc$ take values in $\{1,\ldots,8\}$ only.

As usual for the $\Cwtn$ series, the field strength tensor is defined as 
\begin{equation}
F_{\mu\nu}:=D_\mu D_\nu 1-D_\nu D_\mu 1.\Ptagref{II}{eq:FDD1}
\end{equation}
and a Lagrangian may be constructed as
\begin{equation}
\mscr{L}_\fg=-\frac{1}{4}\Tr{(F^{\mu\nu}F_{\mu\nu})}+\ldots\label{eq:Lfg1}
\end{equation}
where the additional terms are emergent mass terms due to interactions with the pseudovacuum, or involve composite fermions specified later.
Since the tensor product structure ensures that all matrices $\ulambda^A_\tta$ commute with all matrices $\ulambda^C_\ttc$, and both $\ulambda^A_\tta$ and $\ulambda^C_\ttc$ commute with $\ulambda^N$, this expression includes no cross-terms (for example, no terms involving both $A$-type and $C$-type bosons). 
This permits separation of Lagrangian~\eref{eq:Lfg1} by sector to yield
\begin{align}
\begin{split}
\mscr{L}_\fg&=-\frac{1}{4}\Tr{\left(F^{A\,\mu\nu}F^A_{\mu\nu}\right)}-\frac{1}{4}\Tr{\left(F^{C\,\mu\nu}F^C_{\mu\nu}\right)}\\&\,~~~-\frac{1}{4}\Tr{\left(F^{N\,\mu\nu}F^N_{\mu\nu}\right)}+\ldots
\end{split}\label{eq:Lfg2}\\
F^X_{\mu\nu}&=D^X_\mu D^X_\nu 1-D^X_\nu D^X_\mu 1\quad|\quad X\in\{A,C,N\}\\
D^A_{\mu\nu}&=\partial_\mu-\rmi \bmmf\ulambda^A_\tta\fgfield{\bma^{\tta}_\mu}\quad~D^C_{\mu\nu}=\partial_\mu-\rmi \bmmf\ulambda^C_\ttc\fgfield{\bmc^{\ttc}_\mu}\\
D^N_{\mu\nu}&=\partial_\mu-\rmi \bmmf\ulambda^N \fgfield{N_\mu}.
\end{align}
Note that in regimes where the scalar boson can be ignored, the familiar field strength tensor of a Yang-Mills interaction is recovered for $F^B_{\mu\nu}~|~B\in\{A,C\}$:
\begin{equation}
\begin{split}
F^B_{\mu\nu}&:=D_\mu \bmb^{\ttb}_{\nu}\lambda^B_{\ttb}-D_\nu \bmb^{\ttb}_{\mu}\lambda^B_{\ttb}\\
&\p{:}=\,\partial_\mu \bmb^{\ttb}_{\nu} \lambda^B_{\ttb}-\partial_\nu \bmb^{\ttb}_{\mu} \lambda^B_{\ttb}- \rmi \bmmf \left[\bmb^{\ttb}_{\mu} \lambda^B_{\ttb},\bmb^{\ttd}_{\nu} \lambda^A_{\ttd}\right].
\end{split}
\end{equation}

It is also tempting to drop the coloured bosons from the derivative altogether, as was done in \sref{sec:bosonsinn=3}, but care must be taken over the energy scale at which this is done. The $\Cw{18}$ model is seen in \sref{sec:complep1} to contain both colourless and %
weakly coloured composite fermions, with the latter being bound by a residual dipole interaction and thus 
having a binding energy scale no larger than %
$\mc{E}_\preon$. However, it is reasonable to suppose that these coloured emergent particles will still have a characteristic binding energy scale, $\mc{E}_\quark$. Below this energy scale the residual dipole interaction will be yet further reduced, continuing heirarchically until a scale is attained at which the residual force fades to negligible, or at which all observed particles are colour-neutral.

Finally, it is also useful to introduce a notation for complex vector bosons formed from the pairwise recombination of the off-diagonal elements of $\SU{3}_A$. Therefore let
\begin{align}
\left.\bma^{\tta\ttb}_\mu\right|_{\tta<\ttb}&:=\frac{1}{\sqrt{2}}\left(\bma^\tta_\mu + \rmi \bma^\ttb_\mu\right)\label{eq:cplxboson1}\\
\left.\bma^{\ttb\tta}_\mu\right|_{\tta<\ttb}&:=\bma^{\tta\ttb\dagger}_\mu=\frac{1}{\sqrt{2}}\left(\bma^\tta_\mu -\rmi \bma^\ttb_\mu\right).\label{eq:cplxboson2}
\end{align}
For avoidance of ambiguity, complex bosons from the $A$ sector are exclusively described using this notation and not using an $e_{ij}$ representation similar to $\vp^{ij}_\mu e_{ij}$. %

\subsection{Emergent fermions\label{sec:complepE}}

\subsubsection{General construction\label{sec:complep1}}

While the emergent fermions $\psi^{ac}_{\alpha}$ and bosons $\bma^{\tta}_\mu$, $\bmc^{\ttc}_\mu$, $N_\mu$, and $\bmh$ form an effective field theory on $\mbb{R}^{1,3}$ at energies small compared with $\mc{E}_\Omega$, as in \cref{ch:colour} the preons $\psi^{ac\alpha}$ are not observed in the low-energy limit. Instead, they are the constituents %
from which the fermions [and bosons, by \Perefr{I}{eq:opsubvp}{eq:opsubH}]
of the low-energy regime are assembled. 

Consider first the role of the colour charge $c$ on the preons $\psi^{ac}_\alpha$, which is acted on by %
$\SU{3}_C$. Gauge choices in \sref{sec:gaugechoice} leave this symmetry unbroken, and consequently all species carrying colour charges are confined in colour-neutral multiplets. 
In contrast the $\SU{3}_A$ symmetry of the $A$ sector is %
broken in such a way that a persistent charge $a$ may appear on the effective fields of the low-energy regime.

As in \cref{ch:colour}, colour neutrality requires that the effective spinor fields of the low-energy limit are made up of triplets of preons $\psi^{ac}_\alpha$, and requiring the Lagrangian $\mscr{L}$ to have dimension $L^{-4}$ ensures that the only possible spin for an emergent fermion %
appearing in the low-energy limit is $\frac{1}{2}$.
For a general fermion triplet it is therefore convenient to write
\begin{equation}
\begin{split}
\!\!\!\Psi^{\mbf{AC}\alpha}(x_1,x_2,x_3)=\,&\frac{f^2}{\sqrt{3}\mc{N}}\left(\varepsilon^{\alpha\beta}\varepsilon^{\gamma\delta}-\varepsilon^{\alpha\gamma}\varepsilon^{\beta\delta}+\varepsilon^{\alpha\delta}\varepsilon^{\beta\gamma}\right)
\\&\times\psi^{a_1c_1}_\beta(x_1)\psi^{a_2c_2}_\gamma(x_2)\psi^{a_3c_3}_\delta(x_3),
\end{split}{\label{eq:generalfermion}}
\end{equation}
where $\mc{N}$ is a normalisation constant to be fixed later, $\mbf{A}$ is a multi-index short for $a_1a_2a_3$, and $\mbf{C}$ is short for $c_1c_2c_3$. All spinors are \emph{a priori} equally able to interact with other spinors both within and outside the triplet via summation over their spinor index, and to conserve this symmetry there is a normalised summation over the different pairwise contractions of spinor indices within the triplet. 
Let the relevant factor (including associated normalisation),
\begin{equation}
\frac{1}{\sqrt{3}}\left(\varepsilon^{\alpha\beta}\varepsilon^{\gamma\delta}-\varepsilon^{\alpha\gamma}\varepsilon^{\beta\delta}+\varepsilon^{\alpha\delta}\varepsilon^{\beta\gamma}\right),\label{eq:epsilonterms}
\end{equation}
be referred to as the \emph{epsilon terms}.

Note that in any term of the Lagrangian, a factor of $f$ is introduced for %
every particle field after the first two \Perefr{I}{eq:opsubvp}{eq:opsubH}.
When constructing Lagrangian terms for the composite fermions, 
the most convenient normalisation is obtained if
the factor of $f^2$ arising from \PEref{I}{eq:opsubpsi} and the factor of $1/\sqrt{3}$ arising from normalisation of the sum over the epsilon terms~\eref{eq:epsilonterms} %
are absorbed into the definition of $\Psi^{ag\alpha}$ as shown in \Eref{eq:generalfermion} above. 
Also note that the separation of any pair of co-ordinates $x_i$ and $x_j$ in a triplet is at most on order of the preon binding scale.
At lower energy scales it therefore suffices to write $\Psi^{\mbf{AC}\alpha}(x)$ and approximate $x=x_i$ for any $i\in\{1,\ldots,3\}$.
Finally, note that when spinor indices are contracted over a pair of spinors within the triplet, e.g.
\begin{equation}
\varepsilon^{\beta\gamma}\psi^{a_1c_1}_\beta(x_1)\psi^{a_2c_2}_\gamma(x_2),
\end{equation}
such terms do not yield or contribute to a scalar boson $\bmh$ as %
there is not also a sum over the indices $a_i$ and $c_i$ on the spinors $\psi^{a_ic_i}_\alpha(x_i)$.

From the general triplet form~\eref{eq:generalfermion}, three distinct constructions arise.
First, when $a_1=a_2=a_3$, let the colour indices be unique. Symmetry with respect to the indices $a_i$ on $\SU{3}_A$ implies a (complex weighted) sum over permutations of the colour indices, with these weights being determined by requiring that the resulting fermion be an eigenstate of colour exchange process described in \sref{sec:compfermi}, and hence of the mass interaction to be described in \cref{ch:fermion}. For any choice of $a$ it is seen in \cref{ch:fermion} that there are three such eigenstates, in 1:1 correspondence with the eigenvectors of $K$~\Peref{II}{eq:II:KfES},
all net colour-neutral and therefore satisfying confinement with respect to $\SU{3}_C$. Once again summing over the different contractions of the internal spinor indices, these fermions may be written
\begin{align}
\Psi^{ag\alpha}(x) = \,&\frac{f^2}{\sqrt{3}\mc{N}}\left(\varepsilon^{\alpha\beta}\varepsilon^{\gamma\delta}-\varepsilon^{\alpha\gamma}\varepsilon^{\beta\delta}+\varepsilon^{\alpha\delta}\varepsilon^{\beta\gamma}\right)\label{eq:compositeleptonspre}
\\&\times
\mc{C}^g_{c_1c_2c_3}(a)\,\psi^{ac_1}_\beta(x_1)\psi^{ac_2}_\gamma(x_2)\psi^{ac_3}_\delta(x_3)\nn
\end{align}
where $g$ enumerates the three eigenstates which %
correspond to particle generation. The notation $\mc{C}^g_{c_1c_2c_3}(a)$ indicates that the coefficients $\mc{C}^g_{c_1c_2c_3}$ may in general be dependent on particle species $a$. These fermions have no net colour charge, and are identified with the leptons. Writing $\Psi^{ag\alpha}$ in place of $\Psi^{aaag\alpha}$ is a slight abuse of notation, %
but is remedied in \srefs{sec:compbosint}{sec:catalogueall}.

For the second construction, when $a_1=a_2\not=a_3$, again let the colour indices be unique. Symmetry over $\SU{3}_A$ is broken, and in \cref{ch:fermion} it is seen %
that the fermionic mass interaction is dependent on preon charge $a_i$, with equivalent implication for preon mass. In \sref{sec:quarksgluons} it is seen that choices of gauge outlined in \sref{sec:gaugechoice} restrict the permissible triplets of this form, and from the mechanism of the mass interaction discussed in \cref{ch:fermion} (revisited in \cref{ch:detail} and \sref{sec:neutrinos}) it follows that preon~3 is of different mass to preons~1 and~2. Since preon~3 is subject to the same binding force as preons~1 and~2, its greater or lesser inertia contributes to greater or lesser spatial excursions relative to the centre of mass of the triplet, which consequently exhibits a net colour dipole.
Shielding effects in $\SU{3}_C$ are anticipated to mask the inner regions, resulting in a residual colour interaction not stronger than the preon binding interaction (and consequently with characteristic energy scale $\mc{E}_\quark<\mc{E}_\preon$), with the three choices of colour on $a_3$ yielding three different colours of composite fermion. Again, the eigenvalues of the mass interaction yield a generation structure. 
These fermions may be conveniently written in the form
\begin{align}
\begin{split}
\Psi^{a_1a_3g\alpha}(x) = &\frac{f^2}{\sqrt{3}\mc{N}}\left(\varepsilon^{\alpha\beta}\varepsilon^{\gamma\delta}-\varepsilon^{\alpha\gamma}\varepsilon^{\beta\delta}+\varepsilon^{\alpha\delta}\varepsilon^{\beta\gamma}\right)
\\&\times
\mc{C}^{g}_{c_1c_2c_3}(a_1,a_3)\\&\times\psi^{a_1c_1}_\beta(x_1)\psi^{a_1c_2}_\gamma(x_2)\psi^{a_3c_3}_\delta(x_3).
\end{split}
\label{eq:compositequarkspre}
\end{align}
Those not eliminated by gauge correspond to the quarks, with the full set of these fermions being enumerated in \sref{sec:quarksgluons}. As a matter of convention, when a residual colour charge exists it may optionally be indicated by a bracketed label $^{(c)}$. Spinor $\Psi^{a_1a_3g\alpha}(x)$ may consequently also be denoted $\Psi^{a_1a_3(c_3)g\alpha}(x)$

Third, when all indices $a_i$ are unique, satisfaction of confinement is again approximate up to a possible residual colour charge,
\begin{align}
\Psi^{g\alpha}(x) = &\frac{f^2}{\sqrt{3}\mc{N}}\left(\varepsilon^{\alpha\beta}\varepsilon^{\gamma\delta}-\varepsilon^{\alpha\gamma}\varepsilon^{\beta\delta}+\varepsilon^{\alpha\delta}\varepsilon^{\beta\gamma}\right)\label{eq:compositesidewayspre}
\\&\times
\mc{C}^{g}_{c_1c_2c_3}\,\psi^{1c_1}_\beta(x_1)\psi^{2c_2}_\gamma(x_2)\psi^{3c_3}_\delta(x_3),\nn
\end{align}
and this case is also discussed in \sref{sec:quarksgluons}%
.

For all these species %
the three preons making up a triplet $\Psi^{\mbf{AC}\alpha}$ are typically spatially separated by a distance on order of 
\begin{equation}
\mc{L}_\preon:={\mc{E}_\preon}^{-1}, \label{eq:Lpreon}
\end{equation}
with the location $x$ of the composite particle being well-approximated by the location of any component $x_i$ in energy regimes $\mc{E}\ll\mc{E}_\preon$. 
As in \cref{ch:colour}, the energy scale $\mc{E}_\preon$ is assumed large such that preons exhibit confinement at all energy scales at least up to the breakdown scale of the analogue model at $\ILO{\mc{E}_\Omega}$.

When a triplet $\Psi^{\mbf{AC}\alpha}$ contains one or more foreground preons, the combination of confinement plus the vanishing of background field correlators over distances greater than $\mc{L}_0$ implies that an entire triplet of foreground preons must be present and spatially bound by the exchange of bosons with colour charge. These are constructed for the high-energy regime as per \PEref{II}{eq:tvp} and thus distinct from (though related to) %
$\bmc^{\tc}_\mu$. 
However, in the effective Lagrangian developed below, %
the decomposition of preons in any triplet $\Psi^{\mbf{AC}\alpha}(x)$ according to
\begin{equation}
\begin{split}
\psi^{\alpha ac}&=\bgfield{\psi^{\alpha ac}}+\fgfield{\psi^{\alpha ac}}
\end{split}
\end{equation}
will be dominated by terms of lowest order in the foreground fields. Individually these expansions of $\Psi^{\mbf{AC}\alpha}$ into foreground and background components of the preon fields need not conserve colour charge in the foreground and background fields term by term, but collectively these charges \emph{are} separately conserved for foreground and background fields, at least over length scales $\mc{L}\gg\mc{L}_0$, as required by construction of the pseudovacuum. %

\subsubsection{Lagrangian terms for emergent fermions\label{sec:compbosint}}

Ignoring for the moment triplets of form~\eref{eq:compositesidewayspre}, which will be eliminated by choice of gauge on $\SU{3}_C$ in \sref{sec:GL18Cgauge}, consider again the composite fermions arising from \Erefr{eq:compositeleptonspre}{eq:compositequarkspre}. Heuristically, these fermions will appear in the Lagrangian in terms with a form akin to
\begin{align}
\mscr{L}_{{\Psi}}&\sim\rmi\bar{{\Psi}}\,\Dslash{\Psi}+\ldots.
\end{align}
Introduce the notation
\begin{align}
\bm{\Psi}^{ag\alpha}(x) = &\frac{f^2}{\sqrt{3}\mc{N}}\left(\varepsilon^{\alpha\beta}\varepsilon^{\gamma\delta}-\varepsilon^{\alpha\gamma}\varepsilon^{\beta\delta}+\varepsilon^{\alpha\delta}\varepsilon^{\beta\gamma}\right)\label{eq:collectivecompositefermions}
\\&\times\left(\delta^a_{a_1}\delta_{a_2a_3}+\delta^a_{a_2}\delta_{a_1a_3}+\delta^a_{a_3}\delta_{a_1a_2}\right)\nn
\\&\times
\mc{C}^{g}_{c_1c_2c_3}(a)\,\psi^{a_1c_1}_\beta(x_1)\psi^{a_2c_2}_\gamma(x_2)\psi^{a_3c_3}_\delta(x_3),\nn
\end{align}
and let %
\begin{equation}
\left(\delta^a_{a_1}\delta_{a_2a_3}+\delta^a_{a_2}\delta_{a_1a_3}+\delta^a_{a_3}\delta_{a_1a_2}\right)\label{eq:deltaterms}
\end{equation}
be referred to as the \emph{delta terms}. Using this notation, the set of all Lagrangian terms involving fermions~\erefr{eq:compositeleptonspre}{eq:compositequarkspre} may be written explicitly as
\begin{align}
\mscr{L}_{\bm{\Psi}}&=\rmi\,\delta_{\dot aa}\delta_{\dot gg}\;\bar{\bm{\Psi}}^{\dot a\dot g}\,\Dslash\bm{\Psi}^{ag}+\ldots.
\label{eq:LPsibm}
\end{align}
To see that this does indeed yield all such terms, initially consider the third delta term $\delta^a_{a_3}\delta_{a_1a_2}$. The indices $a_1$ and $a_2$ are summed over, and the term for which $a_1=a$ yields a fermion having the form of \Eref{eq:compositeleptonspre}, while the other two terms yield fermions having the form of \Eref{eq:compositequarkspre}. Explicitly, the resulting $A$-charge triplets are given in \tref{tab:Achargetriples}.
\begin{table}
\begin{center}
\begin{tabular}{ccc}
$a_1$&$a_2$&$a_3$\\
\hline
1&1&1\\
2&2&1\\
3&3&1
\end{tabular}\qquad
\begin{tabular}{ccc}
$a_1$&$a_2$&$a_3$\\
\hline
1&1&2\\
2&2&2\\
3&3&2
\end{tabular}\qquad
\begin{tabular}{ccc}
$a_1$&$a_2$&$a_3$\\
\hline
1&1&3\\
2&2&3\\
3&3&\p{.}3.
\end{tabular}
\end{center}
\caption{Charge triples arising from the third of the delta terms~\peref{eq:deltaterms}. Charge labels \protect{$a_i$} correspond to the \protect{$A$}-charges of the preons in \pEref{eq:collectivecompositefermions}.\label{tab:Achargetriples}}
\end{table}%
The other two delta terms $\delta^a_{a_2}\delta_{a_1a_3}$ and $\delta^a_{a_1}\delta_{a_2a_3}$ then symmetrise this construction over internal redistributions of the $A$-charges.
Term-by-term expansion over $A$-charges in \Eref{eq:LPsibm} therefore exhaustively enumerates all composite fermions~\erefr{eq:compositeleptonspre}{eq:compositequarkspre} with all internal permutations of their constituent charges. These permutations are summed without rescaling, as each represents a separate labelling of the constituent spinor operators and thus physical distribution of the $A$- and $C$-charges. This may be contrasted with the epsilon terms~\eref{eq:epsilonterms}, which represent (different couplings of) a single charge labelling and thus attract a normalisation factor of $1/\sqrt{3}$.

It is tempting to try to identify each object $\bm{\Psi}^{ag\alpha}$ as a single particle species. %
However, the breaking of $\SU{3}_A$ symmetry in the gauge choices of \sref{sec:GL18Cgauge} results in different preon species engaging in differing mass-generating interactions (\cref{ch:fermion}) and distinguishable $A$-boson-mediated force interactions. 
Each set of charge labels in \tref{tab:Achargetriples}
therefore yields a triplet with unique behaviour, propagating according to its own unique on-shell trajectory. Each such triplet
must therefore be viewed as a distinct species, up to redistribution of the $A$-charges within the triplet. %
The choice to subsume both $\mc{N}$ and ${f^2}/{\sqrt{3}}$ into $\bm{\Psi}^{ag\alpha}$ avoids having a leading numerical factor appear on the resulting fermion terms, which take forms resembling %
\begin{equation}
\mscr{L}_{{\Psi}}=\rmi\bar{{\Psi}}^{11}\,\Dslash{\Psi}^{11}+\ldots\label{eq:Lpsi11}.
\end{equation}

\subsection{Choice of gauge\label{sec:gaugechoice}}

\subsubsection{Identification of gaugeable symmetries\label{sec:gaugechoicewhichavailable}}

To specify a choice of gauge on $\Cw{18}$ in the low-energy regime, recognise that the construction of the model thus far has preserved the symmetry group of the underlying manifold, and that this symmetry group may be written
\begin{equation}
\begin{split}
\GL{18}{C}\cong&\p{\otimes}[\SU{3}_A\oplus\GL{1}{R}_A]\\&\otimes[\SU{3}_C\oplus\GL{1}{R}_C]\\&\otimes[\SL{2}{C}\oplus\mbb{C}^\times].
\end{split}\tag{\ref{eq:GL18Cdecomp2}}
\end{equation}
Further, as discussed in \sref{sec:gauge}, the foreground fields of the low-energy limit include emergent gauge bosons 
which promote some of the terms in \Eref{eq:GL18Cdecomp2} from global to local symmetries. %
These local symmetries may then exhibit gaugeable degrees of freedom.

To determine which symmetries are local, it suffices to look at the species appearing in the derivative $D_\mu$~\eref{eq:reducedDwithN}. The generators of $\SU{3}$ commute with the generator of $\GL{1}{R}$, permitting treatment of $\SU{3}_A$ and $\SU{3}_C$ independent of $\GL{1}{R}_A$ and $\GL{1}{R}_C$ [collectively $\GL{1}{R}_N$]. %
The relevant species and associated symmetries are:
\begin{enumerate}
\item $\fgfield{\bma^\tta_\mu}%
$, associated with %
$\SU{3}_A\otimes\bm{1}_C\otimes\SL{2}{C}$,
\item $\fgfield{\bmc^\ttc_\mu}%
$,  associated with $\bm{1}_A\otimes\SU{3}_C\otimes\SL{2}{C}$,
\item $\fgfield{N_\mu}$, associated with $\GL{1}{R}_N\otimes\SL{2}{C}$, 
\item $\fgfield{\bmh}$, associated with $\bm{1}_A\otimes\bm{1}_C\otimes\mbb{C}^\times$, which decomposes as the direct sum of
\begin{enumerate}
\item $\bm{1}_A\otimes\bm{1}_C\otimes\mbb{R}^+$ and
\item $\bm{1}_A\otimes\bm{1}_C\otimes\U{1}$, and
\end{enumerate}
\item In principle, also $(\omega_\mu)_\beta^{\p{\beta}\alpha}$, the space--time connection associated with $\bm{1}_A\otimes\bm{1}_C\otimes\SL{2}{C}$, which vanishes in Cartesian co-ordinates on $\RM$.
\end{enumerate}
The notation $\bm{1}$ denotes the trivial representation of any group, but in these examples, $\bm{1}_B$ specifically relates to a trivial representation of $[\SU{3}\oplus\GL{1}{R}]_B$, $B\in\{A,C\}$.
As in previous chapters, mapping $\G$ is from $M\in\Cw{2\bullet}$ to the flat space--time $\RM$, which fixes the connection $(\omega_\mu)^{\p{\beta}\alpha}_\beta$ up to a choice of co-ordinates. %
Also as previously, the introduction of a pseudovacuum associated with a definite energy scale $\mc{E}_0$ breaks scale invariance, preventing gauging of $\bm{1}_A\otimes\bm{1}_C\otimes\mbb{R}^+$
in the low-energy regime. The residual, gaugeable symmetries are therefore associated with bosons as follows,
\begin{equation}
\begin{split}
\SU{3}_A\otimes\bm{1}_C\otimes\SL{2}{C}\quad&:\quad \{\fgfield{\bma^\tta_\mu}\}\\
\bm{1}_A\otimes\SU{3}_C\otimes\SL{2}{C}\quad&:\quad \{\fgfield{\bmc^\ttc_\mu}\}\\
\GL{1}{R}_N\otimes\SL{2}{C}\quad&:\quad \fgfield{N_\mu}\\
\bm{1}_A\otimes\bm{1}_C\otimes\U{1}\quad&:\quad \fgfield{\bmh},
\end{split}
\end{equation}
with gauge parameters arising from the $\SU{3}_A$, $\SU{3}_C$, $\GL{1}{R}_N$, and $\U{1}$ subgroups only.
As noted above, no gauge freedoms are associated with the $\SL{2}{C}$ symmetry of the tangent space, as the connection on this space is fixed (up to purely co-ordinate-based effects) by the choice that $\G(M)$ is flat. Similarly, the parameter space of the $\bmh$ boson is two-dimensional, but a second gauge parameter would correspond to the broken symmetry $\bm{1}_A\otimes\bm{1}_C\otimes\mbb{R}^+$. Thus only the degree of freedom associated with the $\U{1}$ subgroup of $\mbb{C}^\times$ is gaugeable.

\subsubsection{General considerations for \protect{$\SU{3}_B$}}

First, consider the $\SU{3}_B$ symmetry groups, 
each of %
which is an eight-dimensional Lie group %
acting on a five-dimensional vector space%
. The vector space may be parameterised by
analogues to the Euler angles (\aref{apdx:SU3gauge}), and choice of gauge corresponds to a reorientation of a state vector on the locally associated 5-sphere. The gauge parameter space is therefore also five-dimensional%
. Gauge conditions may be expressed as
linearly independent constraints on either the spinors making up the state vector, or the boson fields making up the representation of the Lie algebra, with these being to some extent interchangeable due to the preonic construction of the boson fields.
Given five such conditions, gauge singularities arise when the values of one or more gauge parameters prevent the other gauge parameters from being chosen such that all conditions are satisfied. To count the maximum dimension of gauge singularities on $\SU{3}$, it is helpful to first consider $\SU{2}$.

The gauge parameters on $\SU{2}$ correspond to the Euler angles parameterising a two-dimensional vector space. These parameters may always be chosen such that they are in 1:1 correspondence with the gauge conditions. Gauge singularities arise when the polar angle $\phi\in[0,\pi]$ is 0 or $\pi$, such that all values of the azimuthal angle $\theta\in[0,4\pi)$ then refer to the same point in parameter space.

As observed in \aref{apdx:SU3gauge}, $\SU{3}$ may be understood as two independent copies of $\SU{2}$, denoted $\SU{2}_A\otimes\SU{2}_B$ in that Appendix, plus a third copy $\SU{3}_C$ for which one gauge degree of freedom is fixed by the choice of gauge on $\SU{2}_A\otimes\SU{2}_B$. The residual degree of freedom is isomorphic to $\U{1}$, and therefore may be described as azimuthal, but is nonsingular. To see this, recognise that fixing the polar angle at zero or pi would eliminate two degrees of freedom, not one, and therefore in a co-ordinate system where the free parameter is identified with the angle of azimuth, the polar angle must be non-extremised. Alternatively, if the co-ordinates are chosen such that the polar angle \emph{is} extremised, the residual $\U{1}$ degree of freedom is not mapped to the azimuthal angle alone, but rather parameterises a closed curve in polar co-ordinates $(\phi,\theta)$.
Thus $\SU{3}$ has two polar degrees of freedom, three azimuthal degrees of freedom, and singularities may be of dimension up to two.

\subsubsection{Freedom to choose singularities on $\SU{3}$\label{sec:choosesymbreaks}}

Any one polar and associated azimuthal degree of freedom together comprise co-ordinates on a region of parameter space which is a double cover of the sphere, and is isomorphic to the gauge parameter space of $\SU{2}$. Given such a pair of co-ordinates, it is always possible to perform a change of basis to choose any axis of the sphere as the polar axis.

Let there exist five gauge conditions on $\SU{3}$. Select two of these conditions which are desired to be satisfied everywhere (without singularity). By considering an appropriately chosen $\SU{2}\otimes\SU{2}$ subgroup of $\SU{3}$, at any point on $\RM$ it is always possible to perform two such reparameterisations such that on gauging, the two chosen conditions are satisfied. %
By local symmetry, such reparameterisations may be performed independently everywhere on $\RM$, allowing choice of which gauge conditions are guaranteed nonsingular and which are potentially singular.

To see explicitly how this arises, consider a region which exhibits failure of two gauge choices $G_1$ and $G_2$ from a linearly independent set $\{G_1,\ldots,G_5\}$.
Let another singular region correspond to failure of some other set of two gauge choices, of which at most one coincides with the first region. Since the maximum dimension of the gauge singularity is two, there must exist a closed boundary on which the only unsatisfied gauge conditions are at most those common to both groups, or else
\begin{itemize}
\item the dimension of the singularity would be greater than two, which is prohibited, or
\item the new set of up to two broken gauge conditions on the boundary describes a further, distinct singular region and this argument may be applied recursively to this new (lower-dimension) region and either or both of the original two. %
\end{itemize}
Thus there always exists such a closed boundary or set of boundaries within which one or more $\SU{3}$ co-ordinate (basis) transformations may be performed to match the singularities within each region with the singularities within any other contiguous singular region, merging the two regions. If there is no other such region, then an arbitrary $\SU{3}$ change of basis may be performed on the region, mapping the singularities to gauge conditions not chosen to be nonsingular.

\subsubsection{Specific considerations for \protect{$\GL{1}{R}$} and \protect{$\U{1}$}\label{sec:T2sym}}

First, consider symmetry group $\GL{1}{R}_N\otimes\SL{2}{C}$. The gaugeable degree of freedom associated with this symmetry group arises from the $\GL{1}{R}_N$ subgroup, and takes a value anywhere in the component of the group which is contiguous with the identity. Including the point at infinity this subgroup corresponds to the completion of $\GL{1}{R}_N^+$, which is isomorphic to $\U{1}$.

Second, consider symmetry group $\bm{1}_A\otimes\bm{1}_C\otimes\U{1}$ which must act homogeneously on all elements of the $A$ or $C$ sectors. Recognising that the decomposition of $\GL{9}{R}$ into $\GL{3}{R}_A\otimes\GL{3}{R}_C$ is not unique, the degree of freedom associated with this symmetry may be chosen to parameterise a $\U{1}$ trajectory within the space of these decompositions, yielding a mixing of the $\fgfield{\bma^\tta_\mu}$ and $\fgfield{\bmc^\ttc_\mu}$ bosons and associated $R$-indices. Intuitively this symmetry may be thought of as introducing a gauge parameter $\theta_{\rmU 1}(x)$ which mixes the $a$ and $c$ indices in some specific co-ordinate frame through the transformation
\begin{equation}
\!\Weyll{\psi^{ac}_\alpha(x)}{\psi^{ca}_\alpha(x)}=\bgridtt \cos\,\theta_{\rmU 1}(x) & \sin\,\theta_{\rmU 1}(x)\\
-\sin\,\theta_{\rmU 1}(x) & \cos\,\theta_{\rmU 1}(x) \egridtt \Weyll{\psi^{ac}_\alpha(x)}{\psi^{ca}_\alpha(x)}\label{eq:protoU1gauge}
\end{equation}
applied to all spinors for which $c\not=a$,
with corresponding redefinitions on the 81-element boson sector of \Eref{eq:reducedD81} and subsequent amended reduction to the 17-element boson sector of \Eref{eq:reducedDwithN}. This is 
supplemented by the performance of global co-ordinate transformations on $\SU{3}_A$ and $\SU{3}_C$ prior to and following mixing, such that this mixing is not constrained to be between identically numbered bosons in some arbitrarily chosen set of bases on the groups $\SU{3}_B~|~B\in\{A,C\}$.

Collectively, the parameter space of choices of gauge on $\GL{1}{R}_N\otimes\SL{2}{C}$ and $\bm{1}_A\otimes\bm{1}_C\otimes\U{1}$ is isomorphic to $T_2$, the co-ordinates on the torus, and hence has no co-ordinate singularities.

\subsubsection{Global choice of co-ordinates on \protect{$\GL{18}{C}$}}

In addition to the choices of gauge described above, 
also note that the entirety of $M\subset\GL{18}{C}$ may be covered by a single, global co-ordinate patch.
The manifold $\GL{18}{C}$ is anisotropic, in the sense that there are no privileged co-ordinates in a faithful represetation of this group, and given an initial choice of co-ordinates on $\GL{18}{C}$, any co-ordinate parameter may freely be mixed or interchanged globally with any other(s) to yield another equivalent patch. [In this sense, examples of anisotropic groups would include $\GL{n}{C}$, $\SU{n}$, and $\SO{n}$, but not $\SO{m,n}$ which distinguishes between spacelike and timelike axes.] This mixing of co-ordinates corresponds to the action of the \emph{global} symmetry $\GL{18}{C}$. Incidentally, note that if the $\bm{1}_A\otimes\bm{1}_C\otimes\U{1}$ symmetry is the first to be gauged, then the existence of the $\GL{18}{C}$ global symmetry guarantees that the parameter space of this local symmetry may be identified with transformation~\eref{eq:protoU1gauge} as described above, by bracketing the gauge operation with appropriate global transformations.

\subsubsection{Specification of gauge conditions\label{sec:GL18Cgauge}}

As the vector bosons associated with local symmetry emerge only at energy scales small compared with $\mc{E}_\Omega$ and $\mc{E}_\preon$, the associated gauge conditions are likewise defined in this lower-energy regime. (In higher-energy regimes these choices of gauge map onto choices of co-ordinate frame.) Choice of gauge proceeds as follows:

First, fix the gauge associated with $\U{1}$. Let this gauge mix the $A$- and $C$-sectors of the model, as described in \sref{sec:T2sym}, and require that
\begin{equation}
\delta_{\dot gg}\delta_{\dot aa}\fgfield{\bar\Psi^{\dot a\dot g}\bsm\Psi^{ag}}\bgfield{N_\mu}=0.\label{eq:U1gauge}
\end{equation}
Recognising that the $g$ index forms a basis for the $\SU{3}_C$ sector, in the presence of any single composite fermion this gauge condition imposes that there is no coupling between the fermion and the background $N_\mu$ field.

To see that this gauge condition may be satisfied, let $\Psi^{ca}_\alpha$ denote the preon triplet obtained on interchanging the values of the $a$- and $c$-indices in $\Psi^{ac}_\alpha$ in some specific co-ordinate frame. %
If $\Psi^{ac}_\alpha$ and $\Psi^{ca}_\alpha$ are linearly independent, a Gram--Schmidt decomposition yields a basis of a 2-D vector space, and there always exists a choice of basis on this space such that \Eref{eq:U1gauge} is satisfied. If $\Psi^{ac}_\alpha$ and $\Psi^{ca}_\alpha$ are not linearly independent, adopt an alternative choice of co-ordinate frame which yields a different spinor $\Psi^{ca}_\alpha$ on exchanging the values of indices $a$ and $c$ \emph{in this frame} (noting that index exchange is defined with respect to the values of the indices {in a specific frame.}) By the nonsingularity of the $T_2$ gauge sector as described in \sref{sec:T2sym}, there always exists a global choice of frame such that \Eref{eq:U1gauge} is satisfied everywhere.

The rationale for this choice of gauge is to ensure vanishing of $N$-boson contributions to lepton mass in \cref{ch:fermion}. Gauge condition~\eref{eq:U1gauge} is a nonsingular gauge condition in the $T_2$ gauge subgroup, satisfied everywhere.

Second, fix the gauge associated with $\GL{1}{R}_N$. In some choice of global co-ordinate patch on $M$, this choice of gauge corresponds to local mixing of the real and imaginary parts of the 18-element complex state vector $\psi^{\bm{\alpha}}\equiv\psi^{n\alpha}$. However, since this is only the second choice of gauge, this may be followed by any global co-ordinate transformation on $\GL{18}{C}$ provided it does not mix the $A$- and $C$-sectors and therefore disrupt gauge choice~\eref{eq:U1gauge}. The choice of gauge under $\GL{1}{R}_N$ may therefore correspond to satisfaction of any gauge condition on $\GL{18}{C}$ which is linearly independent of condition~\eref{eq:U1gauge}. This gauge condition may likewise be conserved through all subsequent gauge choices on other sectors, so long as the subsequent conditions imposed are linearly independent of it. As the second gauge choice, impose the condition
\begin{equation}
\delta_{\dot gg}\delta_{\dot aa}\fgfield{\bar\Psi^{\dot a\dot g}\bsm\Psi^{ag} N_\mu}=0\label{eq:GL1RNgauge}
\end{equation}
which is homogeneous with respect to both $A$-charge and colour.
As a result of this choice, any isolated foreground colourless emergent fermion has vanishing net coupling to the foreground $N_\mu$ field. Its constituent preons may still couple to the field, but only for purpose of mutual interactions having no consequences for the dynamics of the fermion as a whole. In addition, foreground $N$ bosons may still moderate contact interactions between co-located foreground fermions of different species as these need only \emph{collectively} sum to zero. This is consistent with the identification of contact interactions as occurring when the regions of space occupied by two preon triplets overlap, in which context it is natural to expect $N$-mediated interactions between the triplets as well as among preons of a single triplet. 
This also is a nonsingular gauge condition in the $T_2$ gauge subgroup, satisfied everywhere.

Third, to gauge the $\SU{3}_C$ subgroup, let the magnitude of the foreground %
complex boson field $\fgfield{\bma^{12}_\mu}$ be set to zero and consider the associated preon expansion
\begin{align}
\begin{split}
\left\|\fgfield{\bma^{12}_\mu}\right\|^2&=[\bma^{21\mu}\bma^{12}_\mu]_\fg\\
&=\delta_{\dot cc}\fgfield{\bar\psi^{2\dot c}\bsm\psi^{1c}}\ \delta_{\dot c'c'}\fgfield{\bar\psi^{1\dot c'}\bsmm\psi^{2c'}}\\
&=-2\left\|\sum_{c,c'}\fgfield{\psi^{1c}\psi^{2c'}}\right\|\\&=0.
\end{split}\label{eq:Hgaugepre}
\end{align}
Recognise that the bound composite $\fgfield{\bma^{12}_\mu}$ appears pointlike at energy scales $\mc{E}\ll\mc{E}_\preon$, and therefore treat the preons comprising this composite particle as co-located. 

Imposing condition~\eref{eq:Hgaugepre} requires only one gauge degree of freedom. However, in principle a global co-ordinate transformation on $\GL{18}{C}$ may independently redefine colour indices for $a=1$ and $a=2$, and it is therefore more general, and also more useful, to impose this constraint as the result of a collection of subconstraints which ensure it is independent of such a redefinition.
Therefore introduce the concept of \emph{relative} and \emph{absolute} colour transformations. An {absolute} transformation acts on the colour indices independent of the value of $a$,
\begin{equation}
\psi^{ac'}=\mc{T}^{c'}_{\p{c'}c}\psi^{ac},
\end{equation}
whereas a more general colour transformation may exhibit $a$-dependence without mixing the different sectors enumerated by $a$, i.e.{}
\begin{equation}
\psi^{ac'}=\sum_j\delta^{ja}[\mc{S}^{(j)}]^{c'}_{\p{c'}c}\psi^{ac}.
\end{equation}
Any such transformation may be written as the product of an absolute transformation $\mc{T}$ acting on all colour sectors, and {relative} transformations $\mc{R}^{(j)}$ acting on the $a=2$ and $a=3$ sectors,
\begin{align}
\psi^{ac''}&=\sum_j[\mc{R}^{(j)}]^{c''}_{\p{c''}c'}\mc{T}^{c'}_{\p{c'}c}\psi^{ac}\\
\mc{R}^{(j)}&=\mc{S}^{(j)}\mc{T}^{-1}\\
\mc{S}^{(1)}&=\mc{T}\Rightarrow \mc{R}^{(1)}=\mbb{I}.
\end{align}
For each of $j=2$ and $j=3$, $\mc{R}^{(j)}$ is a faithful representation of $\SU{3}$ denoted $\SU{3}_{\mc{R}^{(j)}}$.
To introduce an equivalent of \Eref{eq:Hgaugepre} which is invariant under the relative colour transformations $\mc{R}^{(2)}$, first %
write
\begin{equation}
\fgfield{\psi^{1r}\psi^{2r}}+\fgfield{\psi^{1g}\psi^{2g}}+\fgfield{\psi^{1b}\psi^{2b}}=0,\label{eq:colourgauge1}
\end{equation}
for one gauge degree of freedom, then recognise that the space of equations spanned by acting transformations in $\mc{R}^{(2)}$ on the colour space of the $a=2$ sector in \Eref{eq:colourgauge1} is a five-dimensional vector space acted on by $\SU{3}_{\mc{R}^{(2)}}$. %
To impose the $\SU{3}$ generalisation of \Eref{eq:colourgauge1} therefore requires the use of all five gaugeable degrees of freedom. The existence of
singularities in the gauge parameter space does not break this invariance, as they are in 1:1 correspondence with singularities in an equivalent parameterisation of the space of state vectors which arises from the actions of $\SU{3}_{\mc{R}^{(2)}}$ on a state $\psi^{ac}$.

Of particular note, the $\SU{3}_{\mc{R}^{(2)}}$-invariant space of constraints arising from \Eref{eq:colourgauge1} includes the expressions
\begin{align}
\fgfield{\psi^{1r}\psi^{2g}}+\fgfield{\psi^{1g}\psi^{2b}}+\fgfield{\psi^{1b}\psi^{2r}}&=0\label{eq:colourgauge2}\\
\fgfield{\psi^{1r}\psi^{2b}}+\fgfield{\psi^{1g}\psi^{2r}}+\fgfield{\psi^{1b}\psi^{2g}}&=0\label{eq:colourgauge3}
\end{align}
obtained from \Eref{eq:colourgauge1} by cycling the colours $c$ on the $a=2$ sector only %
through the sequence 
\begin{equation}
r\rightarrow g\rightarrow b\rightarrow r.\label{eq:CCI} %
\end{equation}
Collectively, conditions~\erefr{eq:colourgauge1}{eq:colourgauge3} guarantee the satisfaction of \Eref{eq:Hgaugepre}. The other two gauge conditions may be written as
\begin{align}
\fgfield{\psi^{1r}\psi^{2r}}+\fgfield{\psi^{1g}\psi^{2g}}-\fgfield{\psi^{1b}\psi^{2b}}&=0\label{eq:colourgauge4}\\
\fgfield{\psi^{1r}\psi^{2r}}-\fgfield{\psi^{1g}\psi^{2g}}-\fgfield{\psi^{1b}\psi^{2b}}&=0\label{eq:colourgauge5}
\end{align}
and cause all foreground $a=1,a=2$ pairs to disappear term by term as well as under a colour cycle invariant sum, with the result that such pairs are vanishing at all length scales and not just over scales at which colour cycle invariant sums may be assumed to exist.

Let any condition which is invariant under cycling {all} colours as per \Eref{eq:CCI} be termed \emph{colour cycle invariant}. Let the three colour-cycle-invariant conditions~\erefr{eq:colourgauge1}{eq:colourgauge3} be the three nonsingular gauge conditions on $\SU{3}_C$. 
This completes the gauging of the $\SU{3}_C$ subgroup.

As a corollary of this choice of gauge, recognise that since any vertex must conserve all notions of charge with respect to $\GL{18}{C}$ and thus with respect to $\SU{3}_A$ and $\SU{3}_C$, if $A$-charges~1 and~2 are inbound at a vertex then they must also be outbound from the same vertex. It then follows that not only $\|\fgfield{\bma^{12}_\mu}\|^2$ but in fact any colour-cycle-invariant vertex involving boson $\fgfield{\bma^{12}_\mu}$ must in fact vanish. This functionally eliminates the $\fgfield{\bma^{12}_\mu}$ boson from the emergent model of the low-energy limit. %

Next, to begin the gauging of the $\SU{3}_A$ subgroup, impose the four conditions
\begin{align}
\left\|\bgfield{a^{12}_\mu}\right\|&=0\label{eq:bga12gauge}\\
\left\|\bgfield{a^{45}_\mu}\right\|&=0\label{eq:bga45gauge}\\
\left\|a^{67}_\mu\right\|&=0\label{eq:bga67gauge}\\
\left\|a^8_\mu\right\|&=0\label{eq:bga8gauge}
\end{align}
where a total field $a^\tta_\mu$ is defined %
by the sum
\begin{equation}
a^\tta_\mu=\fgfield{\bma^\tta_\mu}+\bgfield{a^\tta_\mu},
\end{equation}
again exploiting that the locations of foreground field excitations $\fgfield{\bma^\tta_\mu}$ may be treated as well-defined %
in the low-energy limit.
The fifth condition to be imposed on $\SU{3}_A$ corresponds to a requirement that the mass of the $a^3_\mu$ boson vanishes. The relevant mass mechanism will be described in \cref{ch:boson}, but up to some small corrections this constraint takes the form
\begin{equation}
\delta_{\dot gg}(\lambda_3)_{\dot aa}\bgfield{\bar\Psi^{\dot a\dot g}\bsm\Psi^{ag}}\fgfield{\bma^3_\mu}= 0.\label{eq:ma3gauge}
\end{equation}
Conditions~\erefr{eq:bga12gauge}{eq:bga45gauge} and~\eref{eq:ma3gauge} are chosen to be nonsingular.

Note that in the presence of other gauge conditions which enforce equivalent constraints for the foreground $\fgfield{\bma^{12}_\mu}$ or $\fgfield{\bma^{45}_\mu}$ bosons, conditions~\erefr{eq:bga12gauge}{eq:bga45gauge} may be rewritten as
\begin{align}
\left\|{a^{12}_\mu}\right\|&=0\label{eq:totala12gauge}\\
\left\|{a^{45}_\mu}\right\|&=0\label{eq:totala45gauge}.
\end{align}
For $a^{12}_\mu$, the $\SU{3}_C$ gauge choices~\erefr{eq:colourgauge1}{eq:colourgauge3} imply \Eref{eq:Hgaugepre}, which states that $\|\fgfield{\bma^{12}_\mu}\|^2$ vanishes. Expanding $\|a^{12}_\mu\|^2$ as
\begin{equation}
\begin{split}
\left\|a^{12}_\mu\right\|^2 =\;& \left\|\fgfield{\bma^{12}_\mu}\right\|^2 + \fgfield{\bma^{12\mu}}\bgfield{a^{21}_\mu} \\&+ \bgfield{a^{12\mu}}\fgfield{\bma^{21}_\mu} + \left\|\bgfield{a^{12}_\mu}\right\|^2
\end{split}
\end{equation}
the middle two terms make no contribution in the low-energy limit, as this corresponds to an average over length scales greater than $\mc{L}_0$ and the foreground and background components are uncorrelated in this regime. Thus, given \Eref{eq:Hgaugepre}, \Eref{eq:totala12gauge} implies \Eref{eq:bga12gauge}.
The equivalent foreground constraint for $a^{45}_\mu$ is not introduced until \cref{ch:gravity}, and thus in the present Chapter %
the choice of gauge for $a^{45}_\mu$ is specified on the background fields as in \Eref{eq:bga45gauge}.

The consequences of these gauge choices become apparent on evaluating the masses of the emergent fermions in \cref{ch:fermion}, and boson exchange interactions in \sref{sec:interactions}.

Strictly speaking, for any of the conditions~\erefr{eq:bga12gauge}{eq:ma3gauge}, accidental degeneracies (where the values of background fields or total fields coincide at some point $x$) may prevent satisfaction of that condition everywhere, even when the condition is chosen to be nonsingular. However, with local symmetry only emerging in the low-energy limit and an assumption that measurements are performed with respect to some probe scale $\mc{L}_p\gg\mc{L}_0$ it suffices that a condition is satisfied on average over regions of characteristic length scale $\mc{L}_0$ (see also the discussion of probe scales in \Psref{I}{sec:ProbeOmegaScale}). Maximisation of entropy of the background fields implies that the different boson fields are uncorrelated over length scales larger than $\mc{L}_0$, and thus such accidental degeneracies are not sustained over length scales $\mc{L}\gg\mc{L}_0$. They may then be compensated either by other accidental degeneracies or by deliberate choice of gauge such that conditions~\erefr{eq:bga12gauge}{eq:ma3gauge} are effectively satisfied over regions large compared with $\mc{L}_0$ as desired. %

This completes the gauging of the $\SU{3}_A$ subgroup.

\subsubsection{Consequences for the particle spectrum\label{sec:consequences}}

Some important consequences of the above choices of gauge
are as follows.
Regarding the $N$~boson:
\begin{itemize}
\item Since $\lambda_N$ commutes with all $\lambda^A_\tta$ and $\lambda^C_\ttc$, the $N$ boson does not couple to any of the other vector bosons. 
\item By gauge choices~\erefs{eq:U1gauge}{eq:GL1RNgauge}, neither the background nor the foreground $N_\mu$ fields interact with single emergent foreground leptons, or on average with colourless collections of emergent foreground fermions.
\item Interactions involving $N$ bosons are therefore restricted to 
\begin{itemize}
\item[(i)] interactions on the preon scale, with preons which necessarily carry a colour charge, 
\item[(ii)] interactions with the scalar boson, which is insensitive to the presence or absence of $A$- and $C$-charges, and 
\item[(iii)] interactions in which $N$ bosons are simultaneously emitted/absorbed by two co-located emergent fermions of different species, potentially sustaining a nonvanishing field of even powers of $N_\mu$, e.g.~$N_\mu N_\nu$, $N_\mu N_\nu N_\rho N_\sigma$.
\end{itemize}
\item When quarks are introduced in \sref{sec:quarksgluons} they are posited to be nominally colour-neutral, up to a small residual which arises due to variations in confinement radii of the different preon species, in conjunction with more effective charge shielding of preons of smaller confinement radius. Such interactions are not suppressed by gauge choices~\erefs{eq:U1gauge}{eq:GL1RNgauge} as stated, so could yield one further group of interactions for the $N$~boson. However, where a quark is not collocated with another particle, equivalent conditions to \Erefs{eq:U1gauge}{eq:GL1RNgauge} may be imposed which apply to the quark construction rather than the lepton construction. Thus $N_\mu$/quark interactions are once again restricted to contact interactions as per~(iii) above.
\item Since the pair exchange interaction~(iii) is relatively obscure, it is generally convenient to overlook it, and identify $N_\mu$ as the diagonal gluon $c^9_\mu$, having fermion interactions only on the preon scale with species which carry a residual charge in the $C$ sector. Except when engaging in pairwise emission or absorption by fermions as per item~(iii), the diagonal boson %
is therefore 
indistinguishable from a ninth $C$-sector boson, %
effectively enlarging the symmetry of the colour sector (only) on the preon scale to $\SU{3}_C\oplus\GL{1}{R}_C$.
\item %
Prior to introduction of the pseudovacuum and gauging, complexity of the space--time representation $\GL{2}{C}$ permitted free interconversion between $\SU{3}_B\oplus\GL{1}{R}_B$ and $\GL{3}{R}_B$ for both $A$ and $C$ sectors through application of Lemmas~\Peref{II}{Lem:1}, \Peref{II}{Lem:9}, and~\Peref{II}{Lem:4}, as seen when going from \Eref{eq:GL18Cdecomp2pre} to \Eref{eq:GL18Cdecomp2} in \sref{sec:symC18}. However, introduction of the pseudovacuum broke the scaling component of $\mbb{C}^\times\subset\GL{2}{C}$, and choice of gauge on the $T_2$ gauge subgroup~(\ref{eq:U1gauge},\ref{eq:GL1RNgauge}) eliminated the action of the $\GL{1}{R}_N$ symmetry on the $A$ sector, fixing the choice of symmetry of the $A$ sector as $\SU{3}_A$. Likewise, the explicitly $\SU{3}$-invariant nature of the choice of gauge on the $C$-sector enforces selection of representation $\SU{3}_C\oplus\GL{1}{R}_N$. Any subsequent use of $\GL{3}{R}_B$ representations on these sectors is therefore to be understood as being synthesised from underlying $\SU{3}_B$ bosons, plus the $N$ boson if required. %
In \srefs{sec:photonint}{sec:weakint} this consideration is seen to have %
bearing on
symmetry factors arising from %
the FSFs during boson exchange.
\item While the notation of \Erefr{eq:cplxboson1}{eq:cplxboson2} is preferred for the $A$ sector, the use of colour labels on the $C$ sector favours the $e_{ij}$ representation for complex bosons, often denoted $e^C_{ij}$, e.g.~$\fgfield{\bmc^{rg}_\mu e^C_{rg}}$. These bosons are consequently always to be understood as representing a weighted sum in the manner of \Erefr{eq:cplxboson1}{eq:cplxboson2}, but indexed by the ${r,g,b}$ co-ordinates of the nonvanishing entry in the resulting representation matrix rather than the Gell-Mann matrices \Perefr{II}{eq:Cbasis1}{eq:Cbasis2} used in their construction.
\end{itemize}
Regarding choice of gauge on $\SU{3}_C$:
\begin{itemize}
\item \Eqrefr{eq:colourgauge1}{eq:colourgauge3} imply that colour-cycle-invariant symmetrisations of $\psi^{1c_1}\psi^{2c_2}$ vanish
independently for background and foreground fields, and likewise for $\bar\psi^{1\dot c_1}\bar\psi^{2\dot c_2}$. Fierz identities then imply that any vertex simultaneously incorporating a preon of type~$a=1$ and a preon of type~$a=2$ must likewise vanish. As noted above this immediately and entirely eliminates the $a^{12}_\mu$ boson from the model, independently in both foreground and background fields.
\item The preon triplets of the $\Cw{18}$ model with mixed $A$-sector charges are enumerated and named in \sref{sec:quarksgluons}; 
where such a triplet incorporates a preon with charge~$a=1$ and a preon with charge~$a=2$, the choice of gauge on $\SU{3}_C$ again entirely eliminates any vertices (including propagation vertices) involving this species, comprehensively eliminating all such species from the model.
\end{itemize}

Regarding the choice of gauge on $\SU{3}_A$: %
\begin{itemize}
\item Constraints~\erefr{eq:bga12gauge}{eq:bga45gauge} are inconsistent with the construction criteria of the background fields (\srefr{sec:QLintro}{sec:fgbg}), unless the number of bosons of the relevant species participating in the background fields is zero.
\item In the absence of foreground bosons, the same is true of constraints~\erefr{eq:bga67gauge}{eq:bga8gauge}. 
\item If a foreground boson of types $\fgfield{\bma^{67\bdag}_\mu}$ or $\fgfield{\bma^8_\mu}$ is present, gauge choices~\erefr{eq:bga67gauge}{eq:bga8gauge} must be satisfied by the foreground and background fields collectively. In \sref{sec:vecbosonmasses} it is shown that these bosons are massive, and therefore this constraint cannot be satisfied on-shell by foreground fields alone. At any given point the background fields may be set to complement the local foreground fields such that constraints~\erefr{eq:bga67gauge}{eq:bga8gauge} are satisfied, but this implies correlation between the foreground and background fields which must be sustained over the duration of existence of the foreground excitation. This is prohibited by definition of the background fields, and consequently the presence of an on-shell foreground $\fgfield{\bma^{67\bdag}_\mu}$ or $\fgfield{\bma^8_\mu}$ excitation must be associated with a singularity of the corresponding gauge condition.
\item When there exists a singularity of gauge conditions~\erefr{eq:bga67gauge}{eq:bga8gauge}, if there is no foreground particle of the relevant species, say $a^{67}_\mu$, then:
\begin{itemize}
\item The background $a^{67}_\mu$ field satisfies the same criteria as the background $a^3_\mu$ field, and thus mixing of $a^{67}_\mu$ and $a^3_\mu$ may be performed without jepoardising gauge condition~\eref{eq:ma3gauge} in the low-energy limit. 
\item Maximisation of entropy implies no correlations between the different boson fields, such that this mixing may in general lead to satisfaction of gauge condition~\eref{eq:bga67gauge}, and no singularity.
\item This remains true on introducing small corrections to \Eref{eq:ma3gauge} in \sref{sec:Wmass5v}. %
\item Consequently, where there is a singularity of gauge condition~\eref{eq:bga67gauge} or \eref{eq:bga8gauge} there \emph{must} also be present a foreground boson of relevant species.
\item Thus for these species, foreground bosons and gauge singularities are in %
correspondence. (They are co-incident.)
\end{itemize}
\item When there exists a singularity of gauge conditions~\erefr{eq:bga67gauge}{eq:bga8gauge}, maximisation of entropy of the background fields implies that the background $a^{67\bdag}_\mu$ and $a^8_\mu$ fields admit the same description in terms of $N_0$ and $\omega_0$ as do background fields unaffected by these choices of gauge, e.g.~$a^3_\mu$ or any $c^\tc_\mu$, i.e.{}(as per \sref{sec:operators}) %
\begin{equation}
\quad~~~~\la\left\|\bgfield{a^{67}_\mu}\right\|^2\ra=\la\left\|\bgfield{a^8_\mu}\right\|^2\ra=-{\mc{E}_0}^2=-{N_0}^2{\omega_0}^2.
\end{equation}
\item Constraint~\eref{eq:bga45gauge} applies only to the background fields, so on-shell and off-shell foreground $\fgfield{\bma^{45\bdag}_\mu}$ bosons may be emitted as per usual. However, it is seen in \sref{sec:interactions} that the absence of $a^{45\bdag}_\mu$ bosons from the background fields dramatically reduces the strength of any interactions involving foreground bosons of these types.
\item When considering the interactions of foreground objects carrying a charge only in $\SU{3}_A$, boson exchange involves only species $\fgfield{\bma^{\ta}_\mu}$, with all $\fgfield{\bmc^{\tc}_\mu}$ %
vanishing. Choices of gauge~\erefr{eq:colourgauge1}{eq:colourgauge3} and~\eref{eq:bga12gauge}
further eliminate all %
bosons of type $a^{12\bdag}_\mu$, and gauge choice~\eref{eq:bga45gauge} effectively eliminates background bosons of type ${a^{45\bdag}_\mu}$ over scales greater than $\mc{L}_0$. In %
\cref{ch:gravity} additional degrees of freedom are recruited to eliminate foreground bosons of types $a^{45\bdag}_\mu$. When $\SU{3}_A$ is considered as a mediator of interactions, these changes collectively corresponds to breaking of symmetry down to $\SU{2}\otimes\mrm{U}(1)$ for the %
particles of the low-energy limit. Consequently at energies small compared to $\mc{E}_\preon$ there appears to be no confinement %
associated with symmetry group $\SU{3}_A$. %
It is an open question whether this confinement is truly broken or whether it continues to act through the additional degrees of freedom recruited in \sref{sec:elimGGpairs} %
to eliminate $\fgfield{\bma^{45\bdag}_\mu}$, but these degrees of freedom may be neglected on normal experimental scales and thus the symmetry $\SU{3}_A$ is functionally broken down to $\SU{2}\otimes\U{1}$ with concommittant effective loss of confinement for particles carrying only $A$-charges.
\end{itemize}

Also note that: 
\begin{itemize}
\item Zeroing one component of the pseudovacuum as per, for example, \Erefr{eq:bga12gauge}{eq:bga45gauge} does not increase the magnitude of the other components of the pseudovacuum. This is because the r.h.s. of \begin{equation}
\la\left\|\bgfield{\vp_\mu(x)}\right\|^2\ra_{\mc{L}_0}
= -{\mc{E}_0}^2\label{eq:bgE0} %
\end{equation} 
[based on \PEref{I}{eq:E0} %
with $\vp_\mu$ being the relevant vector boson, e.g.~$a^{12}_\mu$, and $h=c=1$] is not a vector on $\SU{3}$ to be rotated out of alignment with a particular basis element, but rather represents the mean of a stochastic quantity evaluated with respect to a particular element of the co-ordinate frame on the $\SU{3}$ bundle. 
Over any given region, there is no correlation between the co-ordinate transform required to ensure destructive summation of this quantity and the value of an equivalent expectation value on any other boson field. Adoption of a co-ordinate frame which zeroes \Eref{eq:bgE0} for e.g.~$a^{12}_\mu$ therefore has no net effect on the value of \Eref{eq:bgE0} for any other species, and thus setting (for example)
\begin{equation}
\la\bgfield{a^{12\mu}(x)a^{21}_\mu(x)}\ra_{\mc{L}_0}= 0
\end{equation}
does not affect the value of
\begin{equation}
\la\bgfield{a^{3\mu}(x)a^3_\mu(x)}\ra_{\mc{L}_0}= -{\mc{E}_0}^2.
\end{equation}
Further, where transformations in $\SU{3}_A$ or $\SU{3}_C$ are used to zero expectation values of the form of \Eref{eq:bgE0}, these transformations also have no effect on $\la\bgfield{\bmh\bmh^*}\ra$ because the representation of $\SU{3}_A\otimes\SU{3}_C$ associated with $\bmh$ is trivial.

\item Although (colour-neutral) fermions may not exhibit a net coupling to the $N_\mu$ boson, other bosons and preons are permitted to do so. All colour interactions arise at the preon level, and thus the $N_\mu$ boson behaves as a ninth gluon.
\item On the model as a whole, including the Lagrangian, the formulation of the background fields, and the symmetry-breaking processes of \sref{sec:GL18Cgauge}, global $\SU{3}_C$ invariance is conserved. In contrast with $\SU{3}_A$ there consequently exists no global point of reference in colour space, and the three colour charges may freely be globally and locally mixed by transformations in $\SU{3}_C$. However, this does not prevent choosing a local reference point for colour, whether this be defined arbitrarily or by the colour of a specific set of particles. In this context, superselection of colour charge amounts to a requirement that all terms of a superposition must have equal overall colour charge with respect to that reference. As always, charges (including colour charges) may redistribute freely across the unobserved intermediate particles of an interaction. %
\item The effective $\SU{3}\oplus\GL{1}{R}$ symmetry of the $C$~sector is unbroken at the preon scale, with the consequence that gluon interactions at this scale (though not at the quark scale) always involve an equal superposition of all nine $C$-sector boson species (counting the $N$~boson as the ninth gluon). Although interactions involving only pseudovacuum fields are normalised out of foreground calculations (\Psref{I}{sec:normWrtBgFields}), such interactions nevertheless still take place between the short-lived fluctuations of the background field, and as a consequence, the $C$~sector is comprehensively self-homogenising. Thus pseudovacuum expectation values on the gluons satisfy
\begin{equation}
\la\bgfield{c^{\tc\mu}(x)c^\td_\mu(x)}\ra_{\mc{L}_0}= -{\mc{E}_0}^2\label{eq:<cc>}
\end{equation}
regardless of whether or not the gluons form a conjugate pair.
\item Although the emergent model only strictly takes the form of a locally symmetric quantum field theory sufficiently far below energy scale $\mc{E}_0$, the choices of gauge are realised as choices of co-ordinate frame on $\Cw{18}$, which are well-defined at all energy scales. Where these choices of gauge admit a local extrapolation, for example noting that vanishing of $\|\bgfield{a^{12}_\mu}\|^2$ and of all vertices involving $\fgfield{\bma^{12}_\mu}$ may be realised by a choice which zeroes all vertices involving the total $a^{12}_\mu$ field everywhere, this may be enforced equally well at all energy scales, and not just at $\mc{E}\ll\mc{E}_0$. 
\end{itemize}

Note that the above discussion of gauging exploits the separability in the continuum limit of the $A$ and $C$ sectors discussed in \sref{sec:symC18}. %
Explicit extension of these gauge choices to $\GL{9}{R}$ is presented in \aref{apdx:gaugeSU9}.

\subsection{Catalogue of observed species\label{sec:catalogueall}}

\subsubsection{Composite leptons and electroweak bosons\label{sec:catalogue}}

Having constructed the %
emergent leptons (colourless fermions) and long-range bosons of the low-energy limit, it is now useful to identify what roles these will play in the emerging analogue of the Standard Model. %
To this end, make the following identifications. First, for the composite leptons $\Psi^{ag\alpha}$, assign
\begin{align}
\begin{tabular}{lll}
$\Psi^{11\alpha}\rightarrow {\bar{e}_R}^\alpha\qquad$ & $\Psi^{12\alpha}\rightarrow {\bar{\mu}_R}^\alpha\qquad$ & $\Psi^{13\alpha}\rightarrow {\bar{\tau}_R}^\alpha$\\
$\Psi^{21\alpha}\rightarrow {e_L}^\alpha\qquad$ & $\Psi^{22\alpha}\rightarrow {\mu_L}^\alpha\qquad$ & $\Psi^{23\alpha}\rightarrow {\tau_L}^\alpha$\\
$\Psi^{31\alpha}\rightarrow {\nu_e}^\alpha\qquad$ & $\Psi^{32\alpha}\rightarrow {\nu_\mu}^\alpha\qquad$ & $\Psi^{33\alpha}\rightarrow {\nu_\tau}^\alpha.$
\end{tabular}
\end{align}
Here $e_L$ denotes the left-helicity electron Weyl spinor, $e_R$ denotes the right-helicity electron Weyl spinor (so $\bar e_R$ is a left-handed particle), and $\nu_e$ denotes the electron neutrino, which is Weyl. (It %
has no right-handed counterpart, but remains distinct from its antiparticle. If writing in the form of a Dirac spinor in the helicity basis, the right-handed doublet is zero. It is massless and does not participate in neutrinoless double beta decay. An effect analogous to neutrino mixing arises due to emergent properties of $W$~boson vertices discussed in \sref{sec:neutrinos}.) %

As per convention the higher generations are denoted
\begin{align}
\bar e_R&:~\bar\mu_R,~\bar\tau_R\nn\\
e_L&:~\mu_L,~\tau_L\\
\nu_e&:~\nu_\mu,~\nu_\tau.\nn
\end{align}

Second, the $A$-sector bosons which are still physically relevant after %
the gauge conditions have been imposed 
are identified (or named) as follows:
\begin{align}
\fgfield{\bma^3_\mu}&\longrightarrow A_\mu\label{eq:subA}\\
\fgfield{\bma^{45}_\mu}&\longrightarrow G^\dagger_\mu\\ %
\fgfield{\bma^{67}_\mu}&\longrightarrow W_\mu\\ %
\fgfield{\bma^8_\mu}&\longrightarrow Z_\mu.\label{eq:subZ}
\end{align}

\subsubsection{Composite quarks and gluons\label{sec:quarksgluons}}

Returning now to the composite fermions
\begin{align}
\begin{split}
\Psi^{a_1a_3g\alpha}(x) = &\frac{f^2}{\sqrt{3}\mc{N}}\left(\varepsilon^{\alpha\beta}\varepsilon^{\gamma\delta}-\varepsilon^{\alpha\gamma}\varepsilon^{\beta\delta}+\varepsilon^{\alpha\delta}\varepsilon^{\beta\gamma}\right)
\\&\times
\mc{C}^{g}_{c_1c_2c_3}(a_1,a_3)\\&\times\psi^{a_1c_1}_\beta(x_1)\psi^{a_1c_2}_\gamma(x_2)\psi^{a_3c_3}_\delta(x_3).
\end{split}
\tag{\ref{eq:compositequarkspre}}
\end{align}
\begin{align}
\Psi^{g\alpha}(x) = &\frac{f^2}{\sqrt{3}\mc{N}}\left(\varepsilon^{\alpha\beta}\varepsilon^{\gamma\delta}-\varepsilon^{\alpha\gamma}\varepsilon^{\beta\delta}+\varepsilon^{\alpha\delta}\varepsilon^{\beta\gamma}\right)\tag{\ref{eq:compositesidewayspre}}
\\&\times
\mc{C}^{g}_{c_1c_2c_3}\,\psi^{1c_1}_\beta(x_1)\psi^{2c_2}_\gamma(x_2)\psi^{3c_3}_\delta(x_3),\nn
\end{align}
the species arising from these equations are enumerated in \tref{tab:quarks}. Note that the components of the internal sum over $A$-charge on two of the preons have been individually enumerated: Different terms of this sum acquire different masses through the mass mechanism described in \cref{ch:fermion}, and thus behave as independent particle species. In the resulting catalogue of particles the up and down quarks are identified by their charge and coupling to the $W$ boson, which is seen from expansion of \Eref{eq:LPsibm} to be governed by the unique preon in the triplet.
The third quark pair (``vanishing'', $v_L$ and $v_R$) and the three-species triplet (``sideways'', $w_Q$)\footnote{Higher generations are %
\protect{$x_Q$}, ``appeal'', and \protect{$p_Q$}, ``peppermint''. T.~Pratchett, \emph{Lords and Ladies} (Gollancz, 1992).} %
contain preons with $A$-charges~1 and~2, and thus disappear in all colour-cycle invariant contexts by choice of gauge~\erefr{eq:colourgauge1}{eq:colourgauge3} on $\SU{3}_C$, as described in \sref{sec:GL18Cgauge}. 
\begin{table}
\begin{center}
\begin{equation*}
\begin{tabular}{c|ccc}
 & ${a}_1$ & ${a}_2$ & $a_3$\\ 
\hline
$u_L$ & 1 & 1 & 3\\ 
$d_L$ & 3 & 3 & 2\\ 
\end{tabular}\qquad
\begin{tabular}{c|ccc}
 & ${a}_1$ & ${a}_2$ & $a_3$\\ 
\hline
$\bar{u}_R$ & 2 & 2 & 3\\ 
$\bar{d}_R$ & 3 & 3 & 1\\ 
\end{tabular}
\end{equation*}
\begin{equation*}
\begin{tabular}{c|ccc}
 & ${a}_1$ & ${a}_2$ & $a_3$\\ 
\hline
$v_L$ & 1 & 1 & 2\\ 
$\bar{v}_R$ & 2 & 2 & 1\\
$w_Q$ & 1 & 2 & 3
\end{tabular}%
\end{equation*}
\end{center}
\caption{\label{tab:quarks}Enumeration of quark species arising from \protect{\Erefr{eq:compositequarkspre}{eq:compositesidewayspre}}. Labels \protect{$a_i$} enumerate the preons \protect{$\psi^{a_ic_i}$}.}
\end{table}

If (electromagnetic) charge is defined as the strength of a particle's coupling to the $A_\mu$ boson, including the background field effects described in \sref{sec:photonint} and for fermions the normalisation factor of $\frac{1}{\sqrt{3}}$ from definition~\eref{eq:generalfermion}, and the coupling of $e_L$ defines a reference charge $-|e|$, then both $u_L$ and $u_R$ are found to have charges $+\frac{2}{3}|e|$ while $d_L$ and $d_R$ have charges $-\frac{1}{3}|e|$. Furthermore, the colour-neutral dyad $\bar u_L^{\dot c}d_L^c|_{\dot c=c}$ %
is seen from \Eref{eq:LPsibm} to interact with the $W$ boson according to
\begin{equation}
\bmmf\,\bar{d}_L^{\dot c}\,\delta_{\dot c c}\bsm u_L^c W_\mu+\textrm{h.c.}\label{eq:Lweakint}
\end{equation}
whereas the right-handed counterpart is mediated by the $G^\bdag_\mu$ bosons and is therefore suppressed relative to \Eref{eq:Lweakint} by a factor found in \sref{sec:interactions} to be ${N_0}^3[1+\ILO{{N_0}^{-1}}]$ which evaluates to approximately $7\times 10^6$~\eref{eq:valueN0}. %
(Further, this interaction is eliminated entirely in \sref{sec:Rwnf}). %

Note that the different species of preons make differing contributions to fermion mass interactions, exhibit different effective preon masses, and display differing radial profiles under colour confinement. Shielding effects then have implications for exposed colour charge (see \sref{sec:strongint}). Species of the form of \Eref{eq:compositequarkspre} necessarily exhibit a residual colour dipole, 
and therefore also interact by exchange of bosons carrying non-trivial charges with respect to $\SU{3}_C$. The eight bosons $c^\ttc_\mu$ may therefore provisionally be identified with the gluons of the Standard Model, with $c^9_\mu$ an additional, novel species.

\subsection{Interactions\label{sec:interactions}}

This Section examines couplings in the emergent analogue to the Standard Model, with particular attention to the vector bosons of the electroweak sector and the emergence of an analogue to the Glashow--Salaam--Weinberg Lagrangian. The role of symmetry factors, including those arising from the FSFs, is explored.

\subsubsection{Electromagnetic\label{sec:photonint}}
\begin{figure}
\includegraphics[width=\linewidth]{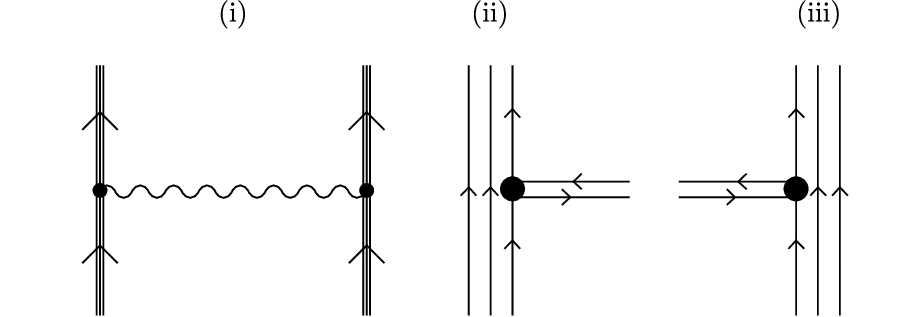}
\caption{(i)~Electromagnetic interaction: Two electrons exchange a photon. Triple lines represent fermions made up of three component preons. (ii)-(iii)~Each interaction vertex is a sum over three terms resembling the ones shown here, with a different preon from the triplet interacting with the photon field in each term. In a lepton the resulting factor of three per vertex cancels with the factor of \prm{\frac{1}{3}} arising from definition~\peref{eq:collectivecompositefermions}.\label{fig:EM}}
\end{figure}

\paragraph[General considerations]{General considerations:\label{sec:photon_general}}
Consider %
the electromagnetic interaction shown in \fref{fig:EM}(i). 
For definiteness, let diagram~(ii) represent emission and diagram~(iii) represent absorption. When a specific preon emits a photon, the raw factor associated with the preon/photon vertex is $f/\sqrt{2}$. Any of the three preons may engage in such an interaction, but the fermion as a whole also incorporates a factor of $\frac{1}{3}$ such that the coupling of the fermion to the photon field attracts an effective vertex factor equal to the mean of the individual preon couplings. That is, a preon with vertex factor (for example) $f/\sqrt{2}$ contributes a term $f/(3\sqrt{2})$ to the vertex factor of the fermion as a whole.

The observed interaction strength is then further modified as emission of a foreground photon always takes place in the presence of the ubiquitous %
background fields. For the fermion this then yields an effective coupling constant of $\bmmf/\sqrt{2}$---in terms of the amended coupling constant $\bmmf$ introduced in \Eref{eq:reducedD81}, and with value now to be determined.

To describe the effect of the background fields on boson emission, recognise that the macroscopic properties of the pseudovacuum may be parameterised in terms of an energy scale $\omega_0$ and a field count $N_0$. In \sref{sec:bgfieldsequiv} it was observed that the background field could under some circumstances be modelled as comprising $N_0$ bosons of each type (or at least, each type not zeroed by gauge), though this interpretation was qualified by the recognition that the degeneracy is in fact not in the bosons but in the %
FSFs %
of the microscopic model. Each time a background vector field enters an expectation value, a vector derivative operator $\bar\partial^{\dot a\dot c}\bsmm\partial^{ac}$ acts on the product of the FSFs, $\vp(x)$. Each spinor derivative may act on any of the $N_0$ FSFs relevant at $x$ [more accurately, there is a weight function such that the integral of this derivative over all FSFs receives nonvanishing contributions equivalent to $N_0$ FSFs; see the discussion around relaxation of the window approximation~\Peref{I}{eq:window} in \sref{sec:denseregime}], and thus the magnitude-squared of a background vector field such as $\la\|\bgfield{a^3_\mu}\|^2\ra$ na\"\i{}evely scales as ${N_0}^4$ prior to being reduced to ${N_0}^2{\omega_0}^2$ through the application of constraints on the background fields.
As noted in \sref{sec:bosonsinn=3}, the constraints which enable this reduction for the background fields do not apply to composite foreground fields.

A key observation in the above is that the symmetry factor of $\ILO{{N_0}^4}$ arises not from interchange of derivative operators, but from interchange of the scalar fields on which they act. With the scalar and vector bosons having equivalent exchange symmetries, in expressions involving bosons it might at first glance seem to make little difference whether one describes the pseudovacuum as $N_0$ photons of mean energy $\omega_0$ within a region characterised by length scale $\mc{L}_0=\mc{E}_0^{-1}$, prototypically a hypercube of side length $\mc{L}_0$ in the isotropy frame of the pseudovacuum, or whether one adopts a description of $N_0$ bosonic scalar fields having their centres within the same region, providing an $N_0$-fold degeneracy of ways to construct a photon, with mean resulting energy $\omega_0$. However, the corresponding distinction is highly significant for fermions, where exchange of FSFs under bosonic statistics allows multiplicative factors similar to those in the vector boson sector to be observed. This, in turn, also has implications for the symmetry factors of the foreground bosons, as these are assembled from non-collocated fermionic preons.

Returning to the process at hand, it is necessary to establish the symmetry factors associated with emission and absorption of a photon. First, consider two distinct cases: 
\begin{enumerate}
\item An emitted photon with energy $\mc{E}_\gamma<\mc{E}_0$, and
\item an emitted photon with energy $\mc{E}_{\gamma'}\geq\mc{E}_0$.
\end{enumerate}
Taking the first of these two scenarios, the photon of energy $\mc{E}_\gamma$ is characterised by a length scale $\mc{L}_\gamma>\mc{L}_0$ such that its wavefunction has nonvanishing overlap with all FSFs within a region characterised by length- and timescales $\mc{L}_\gamma$ in the rest frame of its source. This frame is assumed to coincide with the isotropy frame of the pseudovacuum, or be sufficiently close in the sense of \Psref{I}{sec:pushlimits}. 
There will be a number of FSFs with centres within this region, on which the appropriate derivative operator may act to generate a photon sink---specifically $N_\gamma$, on average satisfying $N_\gamma> N_0$.
However, the correlation length of the background field is $\mc{L}_0$ and within this correlated region there are on average only $N_0$ %
field centres. For definiteness and simplicity, adopt the sharp cutoff approximation~\Peref{I}{eq:window} and choose for there to be a correlated region centred around the %
instant of foreground photon emission. This region contains $N_0$ correlated FSF centres. The other $(N_\gamma-N_0)$ FSFs whose centres are covered by the photon wavepacket, and also all FSFs with centres outside the wavepacket, are uncorrelated. When the derivative operators which generate the photon act on the FSFs, correlated contributions are received from the $N_0$ correlated fields with local centres, while the more distant centres in the background field deliver uncorrelated contributions to whatever quantity is being computed,
with an average which vanishes for sufficiently large probe scale $\mc{L}_p$ (again see \Psref{I}{sec:ProbeOmegaScale}).

For scenario two, $\mc{L}_{\gamma'}<\mc{L}_0$. The wavefunction of the photon now has nonvanishing overlap with fewer FSF centres, $N_{\gamma'}<N_0$, but there exist $N_0$ centres within the local correlated region whose fields are all correlated within the region covered by the wavepacket of the photon. Thus in either scenario, when derivative operators act to construct the photon, the relevant number of background FSFs is $N_0$.

These $N_0$ correlated background FSFs are then supplemented by a further two FSFs with centres lying outside the correlation region but within $\mc{L}_\preon$ of one another. On one of these the appropriate holomorphic derivative operator may act to yield a nonvanishing contribution to boson emission, and on the other the antiholomorphic derivative operator. These FSFs reflect the existence of the long-range correlations corresponding to the emission of a foreground photon made up of two foreground preons. (It is also admissible for one FSF to have nonzero holomorphic and antiholomorphic derivatives, taking the role of both additional sources/sinks, and by independence of the holomorphic and antiholomorphic sectors, this does not change anything in the calculations which follow.)

Precisely the same arguments \emph{also} apply to the interacting foreground preons within the emitting fermion---for the preon lines entering and leaving the vertex, each adds one further FSF for which the relevant derivative operator is nonvanishing, adjacent to the local correlated area.

Na\"\i{}evely one might expect that the emission process would then only correspond to foreground photon emission when the photon-sink-generating derivative operators act on the more distant FSFs which were introduced on account of the foreground boson, and perhaps also those added for the foreground fermion. However, as discussed in \Psref{I}{sec:naturefg}, a foreground excitation is actually a collective property distributed across multiple FSFs. If emission of a foreground photon onto one of the FSFs within the local correlated region takes place, then this requires that one of the implicit bosons described by the local correlation of FSF gradients must now be described as propagating out of the region to the more remote correlated FSF. Collectively, the pseudovacuum remains on average unchanged and correlations propagate \emph{on average} with the nominal foreground photon. Thus an interaction correponding to emission of a foreground photon may take place with the boson's preons not only being emitted onto the added FSFs, but also onto any of the background FSFs within the local correlated region.
Again, a similar argument applies to single preons.

This argument establishes that each preon, be it free, in a boson, or in a fermion, is associated with a factor of $\ILO{\mc{N}_0}$. This factor may be established more precisely for specific processes.

\begin{figure}
\includegraphics[width=\linewidth]{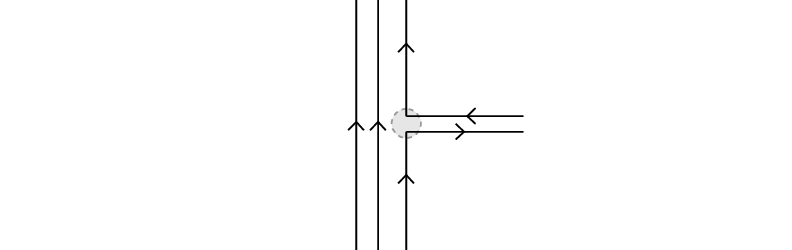}
\caption{In the \prm{\{e_{ij}\}} representation, vertex interactions may be understood as preon-preon scattering (up to a factor in the definition of the fermion which rescales the participating fields, %
discussed further in \psref{sec:scalbosint}).\label{fig:preonpreon}}
\end{figure}

\paragraph[EM symmetry factors]{EM symmetry factors:\label{sec:EWint_sym}} To specifically establish the symmetry factors associated with the emission vertex in \fref{fig:EM}(ii), recognise that in a preon model, using an appropriate representation of $\GL{n}{R}$ for some $n$ [noting comments on $\GL{n}{R}$ in \sref{sec:consequences}], the fundamental interaction process may be conceived as a preon-charge-conserving preon-preon scattering as shown in \fref{fig:preonpreon}.
When, as is required, a basis such as $\{\lambda^A_\ta\}$ is used to represent the action of $\SU{3}_A\oplus\GL{1}{R}_N$ on the $A$-sector, any emission event is always a superposition of all possible events consistent with the inbound and outbound preon lines. Thus at any given instant, without changing species, a preon may (for example) emit a photon, a $Z$ boson, or an $N$ boson, each with some different amplitude accounted for by the different entries in their representation matrices. 
Both in the Standard Model and in the $\Cw{18}$ model, particle masses are ignored during this emission process. In QFT the minimal emission process is represented by a truncated vertex, and in $\Cw{18}$ it is effectively a truncated vertex at length scales $\mc{L}\gg\mc{L}_\preon$.

The upshot of this superposed emission is that coupling to the $A$, $Z$, and $N$ fields is equivalent to coupling to a field $\fgfield{\bm{\vp}^{A\,\dot{a}a}}|_{\dot a=a}$ associated with a representation $e^A_{\dot{a}a}|_{\dot a=a}$ where $a$ is the $A$-charge of the interacting preon. This field in turn may be written as a pair of preons identical to the inbound and outbound legs of the interacting preon in \fref{fig:EM}(ii). All the preons (both in the fermion and in the boson) are constructed by applying spinor derivatives to the FSFs. Counting symmetry factors, there are $N_0+2$ fields capable of supporting a nonvanishing contribution to the first holomorphic derivative, and $N_0+1$ for the second. Likewise, for the antiholomorphic derivatives there are further factors of $(N_0+2)(N_0+1)$ for a total emission symmetry factor of
\begin{equation}
\begin{split}
&(N_0+2)^2(N_0+1)^2 \\&\qquad= {N_0}^4\left[1+6{N_0}^{-1}+13{N_0}^{-2}+\OO{{N_0}^{-3}}\right]
\end{split}\label{eq:EMemitsymfactor}
\end{equation}
\begin{figure}
\includegraphics[width=\linewidth]{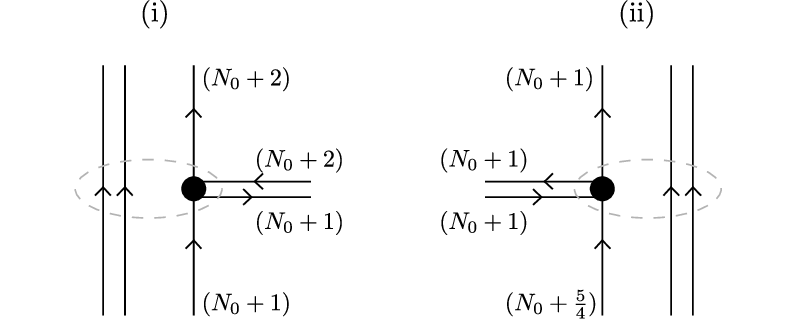}
\caption{Photon emission and absorption redrawn as preon processes. 
The symmetry factors associated with the interacting preons are shown on diagram~(i) for emission and diagram~(ii) for absorption. There is an additional factor of \prm{(N_0+1)^{-2}} on diagram~(ii) corresponding to selection of the foreground preons. These factors are symmetrised to yield the effective vertex factor for the photon, \prm{\bmmf_A}~\peref{eq:EMsymmetrised}. The pair of non-interacting preons in each diagram may also attract symmetry factors, but these are absorbed into a redefinition of the composite fermion field~\peref{eq:collectivecompositefermions}. Their primary consequence is to introduce colour agnosticism at the interaction vertex---if the colours of the interacting preons at the two vertices do not match, they nevertheless acquire the indicated factors regardless, due to the presence of the other preons engaging in the same effective \emph{composite} interaction which is indicated by the dotted ring (and thus located within the same local correlated region).
\label{fig:fermionsymfactors}}
\end{figure}%
as illustrated on \fref{fig:fermionsymfactors}(i). Define
\begin{equation}
\begin{split}
S_{6,13}&:={N_0}^{-4}(N_0+2)^2(N_0+1)^2\\
&\p{:}=\left[1+6{N_0}^{-1}+13{N_0}^{-2}+\OO{{N_0}^{-3}}\right].
\end{split}\label{eq:III:defS613}
\end{equation}

The noninteracting preons in the fermion also attract symmetry factors, and noting that each preon carries a unique pairing of $A$-charge and $C$-charge, the source and sink for each of these preons attracts a factor of $({N_0}+1)$. 
The same factors are present in all fermion figures---interactions, propagators, mass vertices---and thus are absorbed into a redefinition of the fermion field such that the propagator terms in the Lagrangian have no numerical prefactor other than $\rmi$. 
This is achieved through the factor $\mc{N}$ in \Eref{eq:generalfermion} which is now seen to take the value
\begin{equation}
\mc{N}={(N_0+1)^2}.\label{eq:valuemcN}
\end{equation}

For absorption, the process of evaluating the symmetry factors is a little more complex. Recall that as a particle propagates in the presence of the pseudovacuum, it interacts with the gradients of the FSFs. 
Although on average these contributions vanish over the entire trajectory of the photon, %
at the instant when the photon enters the vicinity of the absorption vertex (but before it is absorbed by the preon) it may be in a state transformed by these interactions.

When the photon enters the vicinity of the absorption vertex it does so effectively without accompanying $Z$ or $N$ bosons, as these other species have masses different to that of the photon and thus different on-shell trajectories%
. The engagement of a photon in an interaction vertex is heralded by the appropriate gradients on FSFs outside a local correlated region being correlated with those around a vertex at the centre of that region, similar to the emission process described above. In order for a photon to be present, these correlated gradients must be associated with either or both of the $a=1$ or the $a=2$ charge. On switching to an $e^A_{ij}$ representation to match the preon basis, the probability of both preons in the photon matching both interacting preons in the fermion \emph{at the time that they enter the local correlated region} is $\frac{1}{4}$. 

As with emission, each chiral derivative operator is associated with a symmetry factor of at least $N_0+1$ reflecting the FSFs it may be applied to. However, some of these factors may be increased to $N_0+2$. When the holomorphic derivative in the incoming photon matches the holomorphic derivative in the interacting preon, in principle these admit a factor of $(N_0+2)(N_0+1)$ and not $(N_0+1)^2$. However, the holomorphic and antiholomorphic derivatives on the additional scalar fields are only correlated pairwise (two for the preon and two for the photon) and thus:
\begin{enumerate}
\item If the interacting preon's derivatives are applied first, begin with the holomorphic derivative operator (corresponding to the inbound leg) and recognise it can be applied to any of $(N_0+2)$ FSFs. If applied to one of the added FSFs, the specific counterpart to that FSF is a valid choice for the interacting preon's antiholomorphic derivative operator (outbound leg). However, the foreground FSF from the \emph{other} correlated pair is not correlated with this first pair, and therefore not on average (over many such boson exchanges) capable of yielding a nonvanishing contribution. This reduces the symmetry factor for the antiholomorphic operator to $(N_0+1)$ not $(N_0+2)$.
\item If the holomorphic derivative operator is not applied to an added FSF, the correlation associated with this FSF is still brought into the local area through the action of pseudovacuum fields correlating outside the usual limited area. Again, a specific added FSF is associated with the correlations which arrive with the interacting preon, such that even if the holomorphic operator acts on one of the $N_0$ local FSFs out of the available $N_0+2$, the options available for the antiholomorphic operator are still reduced to $N_0+1$. %
\item This additional freedom of choice for the first foreground operator (chosen, in the above, to be a holomorphic derivative operator) is only possible if both preons in the photon carry $A$-charges matching the interacting preon in the fermion.
\item Connecting the preon lines therefore attracts a factor of $(N_0+2)(N_0+1)$ 25\% of the time, and $(N_0+1)^2$ the other 75\% of the time, for a mean factor of $(N_0+\frac{5}{4})(N_0+1)$.
\end{enumerate}
Further, once the interacting preon's lines are connected, corresponding to making its choices of FSFs, the connection of the photon preons attracts a factor of $(N_0+1)^2$---but for the holomorphic operator only one of the $(N_0+1)$ choices corresponds to a preon in the foreground photon, and similarly for the antiholomorphic operator. If only one preon from the foreground photon is absorbed, this is uncorrelated with the non-foreground counterpart and the resulting vertex vanishes on average over many such interactions. If neither preon from the foreground photon is absorbed, this corresponds to an interaction with an implicit pseudovacuum photon and not with the foreground photon at all. Selection for %
absorption of the foreground photon, completing the foreground EM interaction, therefore corresponds to an additional factor of $(N_0+1)^{-2}$. The net symmetry factor associated with photon absorption is consequently
\begin{equation}
\begin{split}
&\frac{(N_0+\frac{5}{4})(N_0+1)^3}{(N_0+1)^2} \\
&\qquad= {N_0}^2\left[1+\frac{9}{4}{N_0}^{-1}+\frac{5}{4}{N_0}^{-2}+\OO{{N_0}^{-3}}\right].
\end{split}\label{eq:EMabsorbsymfactor}
\end{equation}
Symmetrising across emission and absorption permits calculation of $\bmmf$ for the photon in terms of $f$,
\begin{equation}
\bmmf_A=f{N_0}^3\left[1+\frac{33}{8}{N_0}^{-1}+\frac{687}{128}{N_0}^{-2}+\OO{{N_0}^{-3}}\right],\label{eq:EMsymmetrised}
\end{equation}
and $f$ may be related to $\alpha$. Taking into account higher-order electromagnetic loop corrections (which are equivalent to those of the Standard Model) yields
\begin{align}
\frac{f^2{N_0}^6}{2}&S_\alpha(1\!+\!a_e)^2=\alpha\label{eq:f(alpha)}\\
\begin{split}
S_\alpha:=\,&{N_0}^{-6}{(N_0+\tfrac{5}{4})(N_0+1)^3}{(N_0+2)^2}\\
=\,&\left[1\!+\!\frac{33}{4}{N_0}^{-1}\!+\!\frac{111}{4}{N_0}^{-2}\!+\!\OO{{N_0}^3}\right]
\end{split}\label{eq:defSalpha}
\end{align}
where $a_e$ is the anomaly of the electron magnetic moment (gyromagnetic anomaly of the electron), given to leading order by
\begin{equation}
1+a_e=1+\frac{\alpha}{2\pi}+\OO{\alpha^2}.
\end{equation}
Note that identity~\eref{eq:f(alpha)} implies that $f$, like $\alpha$, has been chosen to subsume the geometric factor of $(4\pi)^{-1}$ associated with emission of a boson field from a point source.

Regarding the gyromagnetic anomaly, it is seen over the course of this and the next few chapters (\ref{ch:fermion}--\ref{ch:detail}) %
that the electromagnetic and electroweak sectors of the $\mbb{C}^{\wedge 18}$ model coincide with the Standard Model to beyond the limit of experimental detection. It is, however, possible that detectable discrepancies may arise as a result of interactions involving the $N$ boson/diagonal gluon, which has no counterpart in the Standard Model.
These effects are not presently anticipated to cross the threshold of detection for the electron, though with this boson having an inertial mass very close to that of the $W$ boson (\srefs{sec:gluonsAndNmass}{sec:results}), %
it is possible the %
effect on the muon gyromagnetic anomaly might be more pronounced. Recent high-precision measurement of the muon magnetic anomaly \cite{aguillard2023} has suggested the existence of tension with the predictions of the Standard Model \cite{aoyama2020}, though the magnitude of this tension is still under debate \cite{borsanyi2021,colangelo2022,ignatov2023}.

\subsubsection{A note on symmetry factors\label{sec:FSFandsym}}

When evaluating more complex diagrams, as will be required in \crefs{ch:detail}{ch:gravity}, it is worth noting that the FSF underpinnings of the $\Cw{18}$ model result in two independent layers of symmetry factors. First, there is the choice of which FSFs support the fermions and bosons of the figure, and then there is the conventional symmetry factor associated with the resulting diagram. Notably, the FSF symmetry factor applies to the scalar fields underpinning fermions as well as bosons.

\subsubsection{Weak sector bosons\label{sec:weakint}}
A similar treatment to the above applies to the exchange of $W^\bdag$ or $Z$ bosons, as the pseudovacuum has nonvanishing $W^\bdag$ and $Z$ components at the site of foreground $W^\bdag$ or $Z$ emission due to the presence of singularities in gauge choices~\erefs{eq:bga67gauge}{eq:bga8gauge} respectively (\sref{sec:consequences}).
This causes the weak sector couplings, like the EM coupling, to be augmented by a factor of $\ILO{{N_0}^6}$. %
Of particular note in deriving \Eref{eq:EMsymmetrised} is $\frac{5}{4}$ term appearing in the construction of \Eref{eq:EMabsorbsymfactor} which arises due to the photon containing both (1,1) and (2,2) entries in its representation matrix $\lambda^A_\ta$. At first glance it might be expected that this effect would not apply to the $W$~boson as it is associated with a single-entry representation matrix. However, as mentioned in \sref{sec:consequences}, gauge choices force adoption of the real bosons of the $\SU{3}_B$ symmetry as physical, and thus the $W$~boson is a convenient description for a superposition of two real bosons, each of which have two nonzero off-diagonal entries in their representation matrices. These real bosons are typically denoted $W^1_\mu$ and $W^2_\mu$ and their actions on $e_L$ and $\nu_e$ are associated with rescalings of the off-diagonal sigma matrices $\sigma_1$ and $\sigma_2$. The $W$ boson therefore attracts the same vertex factor as the photon, $\bmmf_W=\bmmf_A$. 
Calculation of the equivalent factor for the $Z$ boson is not performed explicitly, but it is anticipated to be equivalent by virtue of the underlying $\SU{3}_A$ symmetry.
It is therefore justified to write
\begin{equation}
\bmmf_A=\bmmf_W=\bmmf_Z=:\bmmf.
\end{equation}

\begin{figure}
\includegraphics[width=\linewidth]{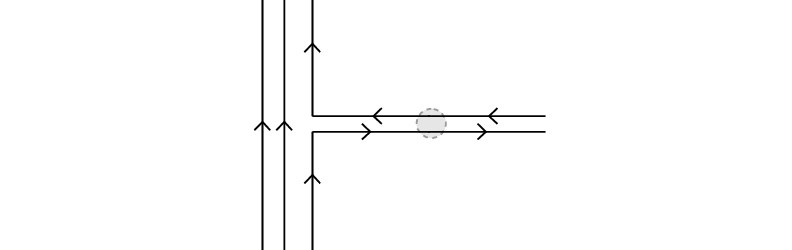}
\caption{In the immediate vicinity of an interaction vertex, a pair of foreground preons (for example, from a fermion as shown here) is indistinguishable from a foreground boson, a qualitative observation which is unaffected by the scaling factor discussed in \psref{sec:scalbosint}. In \protect{\crefs{ch:fermion}{ch:boson}} %
they will be seen to differ only by %
their mass matrices, and thus their on-shell rest energies.\label{fig:preonslikeboson}}
\end{figure}

Consider next the right-handed weak bosons $G^\bdag$. %
As discussed in \sref{sec:consequences}, gauge choice~\eref{eq:bga45gauge} functionally eliminates background $G^\bdag$ bosons. Further, as shown in \fref{fig:preonslikeboson} the foreground preon pair entering the interaction vertex is functionally equivalent to a $G^\bdag$ boson and therefore also does not attract any factors arising from the background field. 
The foreground preons may exchange FSFs, but the overall symmetry factor contains no instances of $N_0$. The resulting expression for $\bmmf_G$ is
\begin{equation}
\bmmf_G=\sqrt{5}f. %
\end{equation}
corresponding to a symmetry factor of $2\times 2\times 1\times 1$ from emission, and from absorption $2\times1\times1\times1$ 25\% of the time and $1\times1\times1\times1$ for the remainder. 
The uninvolved preons are assumed to attract factors of $(N_0+1)$ per source/sink as before.

\subsubsection{\prm{W} interaction and co-ordinate frames\label{sec:EWint_Wintdetail}}
It is now interesting to look at the $W$ boson interaction in more detail, as this is a species-changing interaction for fermions. Considering as an illustrative example the lepton interaction $\fgfield{W_\mu\bar e_L\bsm\nu_e}$, the $W$ boson only couples explicitly to one preon in each of $\bar{e}_L$ and $\nu_e$. Up to a numerical factor, the vertex may be written
\begin{equation}
\left(W_\mu \bar\psi^{2\dot{c}_1}\bsm\psi^{3c_1}\right)\left(\bar\psi^{2\dot{c}_2}\bar\psi^{2\dot{c}_3}\psi^{3c_2}\psi^{3c_3}\right).
\end{equation}
However, recognise that the second factor actually enters the Lagrangian as a sum over species, i.e.{}
\begin{equation}
\left(W_\mu \bar\psi^{2\dot{c}_1}\bsm\psi^{3c_1}\right)\left(\bar\psi^{\p{\dot{a}}\dot{c}_2}_{\dot{a}}\bar\psi^{\dot{a}\dot{c}_3}\psi^{ac_2}\psi^{\p{a}c_3}_{a}\right),
\end{equation}
where the requirement that a well-defined particle must be a definite eigenstate of the mass interaction permits only a single term of this sum to be non-zero in each of the inbound and outbound fermion lines. Further, when the $W^\bdag$ bosons %
interact with quarks, this nonzero term must %
differ for the inbound and outbound fermion lines!
If the $\Cw{18}$ model is to act as an analogue of the Standard Model, it is necessary to also transform the other two outbound preons.

\begin{figure}
\includegraphics[width=\linewidth]{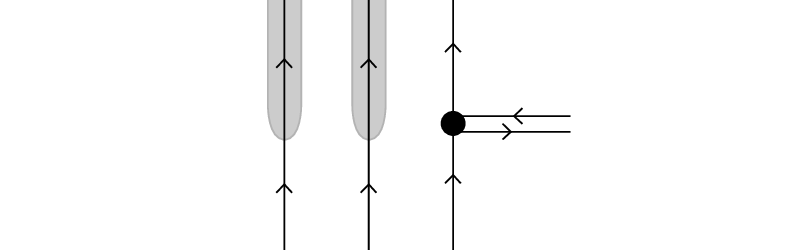}
\caption{The choice of co-ordinates on \protect{$\GL{18}{C}$} is in general predicated on the \protect{$A$}- and \protect{$C$-}charges of a free particle remaining constant over time, except where mandated by fermion/\protect{$W^\bdag$} boson interactions. The ``spectator'' preons, being those in the fermion which do not directly couple with the \protect{$W^\bdag$} boson, are required to undergo a flavour change which is realised by a co-ordinate transformation. In a Feynman path integral approach, this transformation takes place on a hyperellipsoidal boundary just enclosing the affected preon. All preons may be assumed to carry such a co-ordinate ``sleeve'' which is absorbed into the effective definition of the preon.\label{fig:preontube}}
\end{figure}

In the interaction in question, the transformation of the unique preon ($a_3$ in \tref{tab:quarks}) is mediated by a conventional interaction with a gauge boson. However, the mechanism which can be introduced for the paired preons ($a_1$ and $a_2$) is different. %
Given an appropriate inbound fermion and vector boson, the outbound fermion is taken to \emph{define} the necessary outbound species, and this is used to induce an active change of the co-ordinates employed on $\GL{18}{C}$, coincident with the emission event which redefine the $A$-charges. These changes of co-ordinates may be thought of as encompassing open-ended arbitrarily narrow hyperellipsoidal regions of space--time (spherical in spatial cross-section and linear in time) encompassing only the preons undergoing change of flavour, and the resulting flavour-shifting co-ordinate transformation tube is then absorbed into the definition of the preon concerned (\fref{fig:preontube}). 
All preons may then be thought of as being surrounded by such a co-ordinate transformation tube, which may for some be associated with the trivial transformation. In practice such tubes may always be ignored, as (for example) a preon of type~$a=1$ which has acquired a co-ordinate transformation causing it to appear as type~$a=3$ is indistinguishable from any other preon of type~$a=3$. As co-ordinate systems have no independent physical reality, any observer may be assumed to intrinsically adopt a co-ordinate system which ensures simultaneity of the interaction and the co-ordinate change in their own frame of observation. Recognising that the gauge choices of \sref{sec:GL18Cgauge} are likewise identified with nothing more than choices of co-ordinate frame in the underlying model on $\GL{18}{C}$, although this is not a choice of gauge, it is a choice exploiting co-ordinate freedoms on $\GL{18}{C}$ deployed to similar circumstance, namely the description of the emergent model of the low-energy regime using the minimal family of non-prohibited species described in \sref{sec:catalogueall}. %
Ultimately, this choice of co-ordinate frames is compelled by requiring that the $A$-sector bosons act on the fermions as representations of $\SU{2}\otimes\U{1}$, while also remaining consistent with the constraints of gauge (primarily described in \sref{sec:gaugechoice}). This choice of co-ordinate frames is a freedom of description, in much the same way as is a gauge freedom. Other choices of co-ordinate frames permit different descriptions of the same physical system, but a choice in which the bosons act on the fermions as representations of $\SU{2}\otimes\U{1}\otimes\SU{3}_C$ consistent with gauge may always be chosen. The constraints for $\SU{3}_A$ leave only one valid output fermion for any interacting fermion/boson pair.

A more involved form of this scenario is
explored in \cref{ch:detail}. %
In the present scenario, the preon/co-ordinate transformations are independent of the energy scale at which the $W$ boson interaction occurs, whereas in \sref{sec:1storderK}, \emph{energy-dependent} co-ordinate transforms are mapped to interactions involving massless foreground gauge bosons with restricted behaviours. That approach is, however, {only} necessary in situations where the co-ordinate transformation carries an energy/momentum-dependent parameter, and is consequently %
not required for the co-ordinate changes associated with the $W$ boson interactions described here.

More generally, %
there exist three rotation-like subgroups of $\GL{18}{C}$ which may give rise to co-ordinate sleeves of the sort described here, namely $\SU{3}_A$, $\SU{3}_C$, and $\SL{2}{C}$. The equivalent sleeves for $\SU{3}_C$ are fixed in \sref{sec:Csector} by allowing lepton triplets to define colour neutrality%
. %
Note also that by $\SU{3}_C\oplus\GL{1}{R}_N$ invariance of the model at the preon scale, any fermion/gluon interaction is necessarily a superposition of all nine possible gluon interactions. The result is the $K$ matrix of \Psref{II}{sec:compfermi}, explored further in \crefs{ch:fermion}{ch:detail}. %
The choice of frame with respect to the $\SL{2}{C}$ subgroup is described in \cref{ch:gravity}

\subsubsection{Scalar boson\label{sec:scalbosint}}

The electroweak sector is completed by the complex scalar boson $\bmh$, with interactions which may be obtained by expansion of the boson and fermion components of the Lagrangian~(\ref{eq:Lfg2},~\ref{eq:LPsibm}). However, in contrast with the vector bosons, these interactions are generally not augmented by the presence of the background fields. To understand this, it is necessary to examine more closely the manner in which this boson is constructed and contrast this with the vector boson.

The vector and complex scalar bosons are both composed from pairs of preons. In the vector bosons, a pair of preons combine their exposed spinor indices into a single exposed vector index. Each preon comprises a spinor derivative acting on a FSF, and exchanging these FSFs yields different values for the resulting vector boson. On computing the propagator over a distance or time large compared with $\mc{L}_0$, if this propagator is used in the calculation of any physical quantity then the background contributions vanish. For example, if the left-helicity component of the electron field at $x$ exchanges a photon with the left-helicity component of the electron field at $y$,
\begin{equation}
\bar e_L(x)\bsm e_L(x) ~~ A_\mu(x) A_\nu(y) ~~ \bar e_L(y)\bsn e_L(y),
\end{equation}
this contains the photon propagator written in the form $A_\mu(x) A_\nu(y)$, and the background fields make no net contribution to momentum transfer. However, their contribution to $A_\mu(x) A_\nu(y)$ is in general non-vanishing.

This may be contrasted with the scalar boson, which may be written as a sum over nine terms, enumerated by the index pairs $(a,c)$, with internal summation over the space--time indices for each term in this sum. That is,
\begin{equation}
\bmh=f\!\!\!\!\sum_{\substack{a\in\{1,2,3\}\\c\in\{r,g,b\}}}\!\!\!\!\psi^{ac\alpha}\psi^{ac}_\alpha.
\end{equation}
When both spinor derivatives act on the FSFs associated with the longer-range correlations of the foreground field, %
this yields a nonvanishing contribution to $\bmh$. However, if either (or both) acts instead on one of the locally correlated (i.e.~background) FSFs, there is a different %
outcome. The construction of the pseudovacuum guarantees
\begin{equation}
f^2\la\bgfield{\bar\psi\bsm\psi\bar\psi\bsmm\psi}\ra = -2f^2\la\bgfield{\bar\psi\bar\psi\psi\psi}\ra = -{\mc{E}_0}^2\label{eq:bghh*}
\end{equation}
for any spinor $\psi$, but places no constraint on the phase of $\bgfield{\psi\psi}$. 
If the %
preon operators in a pair $\psi\psi$ or $\bar\psi\bar\psi$ arise from spatially or temporally separated vertices connected by a foreground propagator---and they frequently do---%
then by maximisation of entropy in the pseudovacuum it follows that interactions between the intermediate foreground particle and the pseudovacuum randomise the relative phases of the two source/sinks and the phase of the product $\psi\psi$ or $\bar\psi\bar\psi$ is likewise random.
Averaging over multiple such interactions %
involves a sum over phase, yielding zero. %
Similarly, even if both preons arise from the same vertex, if the boson propagates out of the local correlation zone then background-assisted terms pick up phases dependent on but not correlated with the individual preon trajectories. Summation of the background field contributions during propagation therefore does not yield reduction to a classical trajectory, but again leads to a sum over phase, which cancels. 

\begin{figure}
\includegraphics[width=\linewidth]{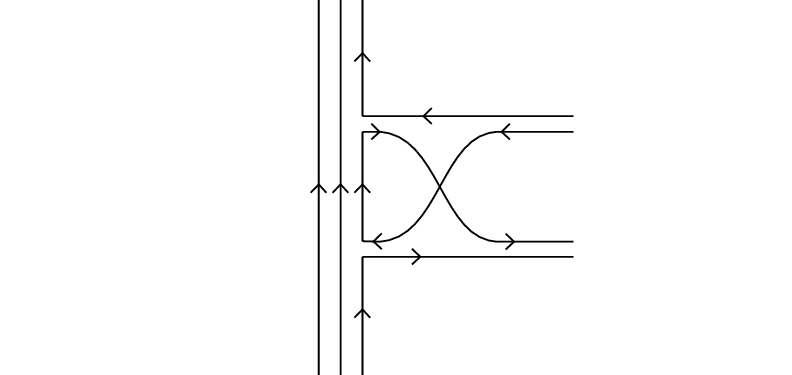} %
\caption{A fermion emits a conjugate pair of complex scalar bosons. Each vertex emits one preon constructed from a holomorphic derivative and one from an antiholomorphic derivative, but these associate pairwise as complex scalar bosons rather than vector bosons. The two interaction vertices are connected by foreground propagation of the fermion. Note that there are two choices regarding which pair of preon lines to cross, but these choices are equivalent up to insertion of a crossing on both far-field complex scalar boson sinks. The sinks incorporate $\varepsilon_{\alpha\beta}$ or $\varepsilon_{\dot\alpha\dot\beta}$ and average over internal configurations, so there is no sign of numerical factor associated with this choice.\label{fig:scalarspread}}
\end{figure}

Regarding the assertion that preon operators in a pair $\psi\psi$ or $\bar\psi\bar\psi$ often arise from spatially or temporally separated vertices,
recall that $\bmh$ and $\bmh^*$ bosons are always constructed pairwise, either from a single vertex or from two vertices separated by no more than $\mc{L}_0$, through the identity
\begin{equation}
\partial_\mu\partial_\nu\equiv-\frac{1}{2}\eta_{\mu\nu}\bar\partial\bar\partial\partial\partial\equiv-\frac{1}{2}\eta_{\mu\nu}\bar\partial_\rmU\partial_\rmU.\tag{\ref{eq:makedU}}
\end{equation}
When the two derivative operators act at spatially distinct sites, the non-chiral nature of the Lagrangian requires that each operator is in fact $\partial_\mu$, and the complex scalar bosons detected in the far field are the result of the way in which the foreground preons assort themselves on leaving the local correlated area, and hence the mass operator for which they are eigenstates (see \cref{ch:boson}). %
This is shown in \fref{fig:scalarspread}. 
The preon propagators reduce the separation of the components of the complex scalar bosons from $\OO{\mc{L}_0}$ to $\OO{\mc{L}_\preon}$, and in this context the loss of background term contributions \emph{always} applies.

In applying the above findings, it is important to take note of the following:
\begin{enumerate}
\item Since the properties of the pseudovacuum do not impose any phase correlations whatsoever, any propagation at all may in theory take a scalar boson into the ``far field'' regime. In practice the relevant scale is likely to be $\mc{L}_\Omega$~\Peref{I}{eq:I:EOmega}, the length scale associated with the highest supported background field energies of the $\Cw{18}$ model.
\item When the preons in \fref{fig:scalarspread} associate to construct foreground scalar bosons in the far field, they are subject \emph{in the far field} to the cancellation described above, for a relative factor of ${N_0}^{-2}[1+\ILO{{N_0}^{-1}}]$ per boson. Since this cancellation is specifically
a function of the way that they assort in the far field, the two preon lines on the composite fermion side of each vertex do {not} behave collectively as a complex scalar boson. They attract uncancelled factors of $\OO{N_0}$ each in the usual way. 
\item This factor is then supplemented by a further factor of~2 corresponding to interchange of the foreground FSFs at the emitting vertex (as both are acted on by nonzero spinor derivatives with the same charge labels). On external scalar boson lines the source/sink is proportional to $\varepsilon_{ab}$ or $\varepsilon^{\dot a\dot b}$, so preon exchange on external lines does not attract a factor of $-1$, and this FSF exchange is equivalent to crossing of the two component preons.
\item Such factors of two for foreground complex scalar boson fields are \emph{only} encountered at interaction vertices absorbing from or emitting to the far field, and not during propagation or at far field sources/sinks. This is because the mass vertex $m_\bmh^2$ and the propagator $\square$, and all proper self energy corrections to the same, act on the internal configuration space of the preon as multiples of the identity. All crossings may therefore always be pushed to one end of a propagator and cancel pairwise. When there exists an external source or sink, this pushing is always to the interaction vertex, though the antisymmetric tensor in the definition of the source/sink ensures addition of terms.
\item In contrast, when a foreground complex scalar boson field has interaction vertices at each end it is in general necessary to explicitly consider both crossed and uncrossed terms. This summation will frequently eliminate diagrams with internal scalar boson lines due to a summation over sign, though examples where it does not may be found in \cref{ch:detail}. %
\item There is a further factor associated with cancellation of the near-field effects. To understand this factor requires tools not developed until \crefr{ch:fermion}{ch:detail}%
, but it is nevertheless presented here for completeness. This factor is denoted $k^{(e)}_1$.

First, recognise that scalar boson interactions must always occur as conjugate pairs within a region of dimension anticipated to be at most of $\ILO{\mc{L}_\Omega}$. Let a test particle engage in such a pair of conjugate interactions with the scalar boson field. Similarly, let another test particle engage in a similar pair of conjugate interactions with a vector boson field. Compare these processes and recognise:
\begin{itemize}
\item As discussed above, there is one arriving boson per interaction vertex, and the holomorphic and antiholomorphic operators at the vertex may each act on any of $N_0$ FSFs. For the vector bosons this yields a factor of ${N_0}^2$ per vertex, but (as discussed above) this factor is absent for the scalar bosons.
\item For vector bosons, the bosons which are absorbed at these vertices are first emitted by one or more excitations of the pseudovacuum state. At the pseudovacuum energy scale $\mc{E}_0$, emitting excitations take the form of composite bosons or composite fermions. 
\item The vertices of this emission process are structurally analogous to the vertices in the mass interactions of \freft{fig:vecbosonmass}(i) and~(iii)-(v). It is seen in \sref{sec:vecbosonmasses} that such couplings are dominated by terms where the pseudovacuum state emitting the vector boson behaves as a lepton [\freft{fig:vecbosonmass}(v)], and it is further seen in \sreft{sec:tooheavy}, \ref{sec:speciesdependenceofk}, and~\ref{sec:bosonkmatrixdependencies}
that 
\begin{itemize}
\item there are three generations of bosons,
\item for first-generation bosons, contributions where the pseudovacuum behaves as \emph{charged} leptons predominate, and
\item the generation of these leptons corresponds to the generation of the boson, here taken to be first-generation.
\end{itemize}
\item It also follows that
\begin{itemize}
\item Where such a pair of vector bosons is emitted by the vacuum state, the associated charged composite lepton interaction vertex {may} attract a pair of additional coefficients arising as eigenvalues of the $K$-matrix of \sref{sec:Csector}. %
As in the boson mass interaction of \sref{sec:Wmass5v}, %
interactions with charged pseudovacuum composite leptons are dominated by terms where these $K$-matrix eigenvalues are indeed present.
The associated factor is denoted $\big[k^{(e)}_1\big]^2$.
\item In the boson mass interactions of \cref{ch:boson}, the factor ${N_0}^2$ is necessarily kept extant for purpose of computing FSF symmetry factors, but the factor of $\big[k^{(e)}_1\big]^2$ is implicitly absorbed within a pseudovacuum mean field term which incorporates all contributions of the pseudovacuum at the interaction vertex except for symmetry factors.
\end{itemize} 
\item None of this applies to the scalar boson. Since the contributions from the pseudovacuum vanish, the only residual term is that in which one vertex acts as the source of the $\bmh$~boson and the other as the sink. The pseudovacuum is identical, but the factors associated with boson emission are lost. Relative to the vector boson, the scalar boson process therefore attracts a factor of $\big[k^{(e)}_1{N_0}\big]^{-2}$, not just ${N_0}^{-2}$.
\end{itemize}
\item For background complex scalar bosons, recognise that although the background fields are typically evaluated according to their mean field values, perturbations around these values may transiently carry momentum borrowed from the foreground field as per \Psref{I}{sec:4momflucs}. To determine the FSF symmetry factors associated with a background vertex, look at the manner in which these perturbations are returned to the foreground fields. If and only if this involves excursion beyond $\mc{L}_\Omega$, there will be a reduction in symmetry factors corresponding to multiplication by $2\big[k^{(e)}_1N_0\big]^{-2}
[1+\OO{{N_0}^{-1}}]$.
\item An example where the reduction factor does not apply is when a pair of complex scalar bosons are emitted by a single vertex, for example in a coupling having the form $V^\mu V^\dagger_\mu\h\h^*$ for some vector boson $V_\mu$, and the scalar boson line is closed in a loop which may then be contracted down to a point. The key features here are that $\bmh$ is both emitted and absorbed on the same vertex, and that the loop may be closed without propagation into the far field regime. 
\end{enumerate}

One net consequence of the phenomena described above is that the long-range coupling constant for the foreground complex scalar interaction admits a correction relative to %
the EM or left-handed weak interactions,
\begin{equation}
\bmmf'=\bmmf \big[k^{(e)}_1N_0\big]^{-2}%
\left[1+\OO{{N_0}^{-1}}\right],\label{eq:fprime}
\end{equation}
where $k^{(e)}_1$ and $N_0$ are evaluated in \sref{sec:results} and \arefr{apdx:solve}{apdx:accessory} as %
$0.04035007804(41)$ and $191.9470(37)$ respectively, giving %
$\bmmf'\approx \frac{1}{60}{\bmmf}$. Note $\bmmf'$ does not include symmetry factors relating to internal crossing as these may be context-dependent. 
The relative weakness of this interaction makes it
unnecessary to explicitly elaborate the higher-order corrections to \Eref{eq:fprime} at this time. 

For interactions over shorter length and timescales, including some interactions with the background scalar field, the presence or absence of the $2\big[k^{(e)}_1N_0\big]^{-2}%
$ scaling factor may depend on whether or not the $\bmh$ and $\bmh^*$ bosons are produced by the same vertex, or merely two separate vertices within the same correlation region. This has been discussed further %
above, with a relevant example in \sref{sec:WmassQLphotonandscalar}.

These points highlight the importance of remaining aware of the preonic underpinnings of the model. In particular, when the interaction vertices are disjoint, the incorporation of $\bmh^{(*)}$ into the covariant derivative as per \PEreft{I}{eq:Dbold} and~\eref{eq:III:expandedD} is a convenient (as opposed to a natural) shorthand for the post-emission assembly of scalar bosons from constituent preons. 

Note that similar relative factors of $\big[k^{(e)}_1N_0\big]^{-2}%
[1+\ILO{{N_0}^{-1}}]$ per source/sink in the far field %
also apply to the mass vertex $\bmh^*m^2_\bmh\bmh$ and the derivative term $\bmh^*\square\bmh$. Both of these Lagrangian terms contribute to the massive propagator, and may be associated with diagrams with untruncated external legs (which are themselves massive propagators). The mass vertex has two scalar bosons propagating into the far field, and the derivative term has identical construction up to a change of diagonal operator from $m^2_\bmh$ to $\square$. By construction these operators are multiples of the identity, and as discussed above, all internal crossings are pushed out of the propagators and only counted at interaction vertices. Consequently the factor associated with taking \emph{foreground} complex scalar bosons (massively) to the far field is $\big[k^{(e)}_1N_0\big]^{-2}%
[1+\ILO{{N_0}^{-1}}]$ and not $2\big[k^{(e)}_1N_0\big]^{-2}%
[1+\ILO{{N_0}^{-1}}]$.

If the Lagrangian is scaled in the usual way, such that there is no prefactor on the derivative terms for off-diagonal vector bosons, and a factor of $\frac{1}{2}$ for diagonal vector bosons i.e.{}
\begin{equation}
\mscr{L}_\mrm{fg} = W^\dagger_\mu(\triangle^{\mu\nu}-m^2_W\delta^{\mu\nu})W_\nu+\frac{1}{2}A_\mu\triangle^{\mu\nu}A_\nu+\ldots,
\end{equation} 
then the scalar boson terms carry additional prefactors %
\begin{equation}
\begin{split}
\mscr{L}_\mrm{fg} = &\ldots -2\big[k^{(e)}_1N_0\big]^{-4}[1+\ILO{{N_0}^{-1}}]\bmh^*(\square-m^2_\bmh)\bmh%
+\ldots
\end{split}\label{eq:Lscalarprop}
\end{equation}
where $-2$ comes from a sigma matrix expansion.
Due to this unusual feature, some caution is %
required when working with complex scalar bosons. It is possible to rescale the $\bmh$ boson to absorb the factor of $2\big[k^{(e)}_1N_0\big]^{-4}$ but not the overall negative sign, and the suggested notation for the rescaled boson is $\tilde{\bmh}$. However, the scaling of \Eref{eq:Lscalarprop} has a major advantage in that the preon fields making up vector and scalar bosons may be interconverted purely by spatial rearrangement as per %
\fref{fig:scalarspread}, making it the natural scaling to employ for calculations approaching the preon scale $\mc{E}_\preon$ at which preons may frequently change their associations. For any other scaling, such a rearrangement is accompanied by %
an additional numerical factor. Note that the fermion fields, on the other hand, \emph{have} been rescaled in this way as described in \srefs{sec:complepE}{sec:photonint}.

As with the Standard Model Higgs boson, the scalar boson is capable of coupling to all other species in the $\Cw{18}$ model. A more comprehensive exploration of the behaviours of the scalar boson is deferred for the time being, though some calculations involving and describing the scalar boson [performed in the natural scaling of \Eref{eq:Lscalarprop}] may be found in \crefr{ch:boson}{ch:detail}. %

\subsubsection{Electroweak Lagrangian\label{sec:EWint_Lagr}}
Consider now the Lagrangian of the electroweak sector as a whole. Defining
\begin{align}
{g}&=\sqrt{2}\bmmf_A\\
{g'}&=\frac{2\bmmf_A}{\sqrt{6}}=\frac{{g}}{\sqrt{3}}
\end{align}
\begin{align}
\tilde{\lambda}^A_\ta&=
\left\{
\begin{array}{lcl}
\lambda^A_\ta&|&\ta\in\{3,6,7,8\}\\
0&|&\ta\in\{1,2\}\\
{2\sqrt{2}}{{N_0}^{-3}}\left[1+\ILO{{N_0}^{-1}}\right]\lambda^A_\ta&|&\ta\in\{4,5\}
\end{array}
\right.
\end{align}
\begin{align}
\Psi&=\triplet{\bar e_R}{e_L}{\nu_e}
\end{align}
and anticipating the mass terms to be described in \crefr{ch:fermion}{ch:boson}, %
an effective Lagrangian may be written down for the interactions of the foreground fields%
, with the role of the background fields being subsumed into the mass terms and the enhanced couplings of the electroweak and coloured bosons. Momentarily restricting attention to just the $A$-sector vector bosons and the first generation of colourless leptons, the foreground Lagrangian~\eref{eq:Lfg2} augmented by the fermion terms of \Eref{eq:LPsibm} takes on a form resembling the corresponding terms of the Glashow--Salaam--Weinberg Lagrangian plus a few additional terms in $G^\bdag_\mu$:
\begin{align}
\begin{split}
\bm{D}_\mu&:=%
\partial_\mu-\rmi \frac{{g}}{\sqrt{2}}\,\fgfield{\bma^\ta_\mu}\tilde{\lambda}_\ta^A\label{eq:covarderiv}
\end{split}
\\
\bm{F^A}_{\mu\nu}&:=\bm{D}_\mu \fgfield{\bma^{\ta}_{\nu}}\tilde{\lambda}^A_{\ta}-\bm{D}_\nu \fgfield{\bma^{\ta}_{\mu}}\tilde{\lambda}^A_{\ta}\\
&\p{:}=\,\partial_\mu \fgfield{\bma^{\ta}_{\nu}} \tilde{\lambda}^A_{\ta}-\partial_\nu \fgfield{\bma^{\ta}_{\mu}} \tilde{\lambda}^A_{\ta}%
- \frac{\rmi {g}}{\sqrt{2}} \left[\fgfield{\bma^{\ta}_{\mu}} \tilde{\lambda}^A_{\ta},\fgfield{\bma^{\tb}_{\nu}} \tilde{\lambda}^A_{\tb}\right]
\nn
\\
\mscr{L}_{\mrm{fg}}&=\mscr{L}_{\not\partial}^{(\mrm{fg})}\!+\!\mscr{L}^{(\mrm{fg})}_{b}\!+\!\mscr{L}_\mrm{GSW}^{(\mrm{fg})}\!+\!\mscr{L}_{G}^{(\mrm{fg})}\!+\!\mscr{L}_{m}^{(\mrm{fg})}%
\label{eq:LGSWfg}\\
\begin{split}
\mscr{L}_{\not\partial}^{(\mrm{fg})}&=\rmi\fgfield{\bar\Psi{\dslash}\Psi}\\
&=\rmi\fgfield{\bar e_R\,{\dslash}\, e_R}+\rmi\fgfield{\bar e_L\,{\dslash}\, e_L}+\rmi\fgfield{\bar\nu_e\,{\dslash}\,\nu_e}
\end{split}
\\
\mscr{L}_{b}^{(\mrm{fg})}&=-\frac{1}{4}\Tr{\left({\bm{F}^{\bm{A}\,\mu\nu}\bm{{F}^A}_{\mu\nu}}\right)}
\\
\begin{split}
\mscr{L}_\mrm{GSW}^{(\mrm{fg})}&=\frac{{g}}{2}\left(\fgfield{\bar\nu_e\bsm\nu_e}-\fgfield{\bar e_L\bsm e_L}\right)\fgfield{W^3_\mu}\\
&\p{=}+\frac{{g}}{\sqrt{2}}\left(\fgfield{\bar e_L\bsm\nu_e W_\mu}+\mrm{h.c.}\right)\\
&\p{=}-{g'}\fgfield{B_\mu}\left(\fgfield{\bar e_R\sigma^\mu e_R}\!+\!\frac{1}{2}\fgfield{\bar e_L\bsm e_L}+\frac{1}{2}\fgfield{\bar\nu_e\bsm\nu_e}\right)
\end{split}
\\
\mscr{L}_\mrm{G}^{(\mrm{fg})}&=\frac{{g}}{\sqrt{2(N_0+1)}}\left(\fgfield{e_R\bsm \nu_e G_\mu^\dagger}+\mrm{h.c.}\right)
\\
\nn\mscr{L}_m^{(\mrm{fg})}&=m^2_W \fgfield{W^\dagger_\mu W^\mu}+m^2_G \fgfield{G^\dagger_\mu G^\mu} + m^2_Z \fgfield{Z_\mu Z^\mu} 
\\
&\p{=}+ \!\!\!\!\!\!\sum_{\ell\in\{e,\mu,\tau\}}\!\!\!\!\!\!m_\ell\left(  \fgfield{\bar \ell_R \ell_L} + \fgfield{\ell_R\bar \ell_L}\right)
\\
\fgfield{A_\mu}&=\frac{{g'}\fgfield{W^3_\mu}+{g}\fgfield{B_\mu}}{({g}^2+{g'}^2)^{\frac{1}{2}}}\\
\fgfield{Z_\mu}&=\frac{{g}\fgfield{W^3_\mu}-{g'}\fgfield{B_\mu}}{({g}^2+{g'}^2)^{\frac{1}{2}}}.
\end{align}
Definition of the $\SU{2}\otimes\U{1}$ boson fields $W^3_\mu$ and $B_\mu$ follows the same convention as \citeauthor{ryder1996}~\cite{ryder1996}. %

Note that:
\begin{itemize}
\item The left- and right-helicity electron spinors can be assembled into the usual Dirac pair. The neutrino is Weyl and massless (see \sref{sec:neutrinos} for neutrino generation mixing in CASMIR).
\item Term $\mscr{L}_\mrm{G}$ provides a form of right-handed weak interaction, with the anti-[left-handed neutrino] serving as the right-handed neutrino. This process is suppressed relative to its left-handed counterpart by a factor of $\ILO{{N_0}^{-3}}$ per vertex due to gauge choice~\eref{eq:bga45gauge} eliminating the FSF symmetry factors associated with the background fields as discussed in \sref{sec:weakint}. This suppression factor is reasonably robust, coming out
at approximately $1.4\times 10^{-7}$ %
when $N_0$ is evaluated as per \arefr{apdx:solve}{apdx:accessory},
making it reasonable to neglect $G^\bdag$ processes pending their complete elimination in \cref{ch:gravity}. %
\item At tree level the weak mixing angle $\tan\theta_W=g'/g$ differs substantially from its assumed value in the Standard Model, as does the mass ratio $m_W^2/m_Z^2$. In the Standard Model these parameters are related through
\begin{equation}
\sin^2\theta_W=1-\frac{m_W^2}{m_Z^2}\qquad\tan\theta_W=\frac{g'}{g},\label{eq:SMthetaW}
\end{equation}
with the lowest-order value of $\sin^2\theta_W$ in the present model being $0.25$, whereas the observed value is \protect{$\sin^2\theta_W=0.22290(30)$~\cite{tiesinga2018}}. 
However, as discussed subsequently in \crefr{ch:boson}{ch:detail}, %
corrections to the mass ratio differ substantially from the Standard Model and preferentially grant mass to the $W$ boson. A more detailed calculation of the $W$:$Z$ boson mass ratio in \sref{sec:weakmix} %
yields a value of $\sin^2\theta_W$ far closer to the Standard Model than the initial %
tree-level approximation %
presented here.
\item These corrections also modify the observed values of the coupling constants $g$ and $g'$ consistent with their effect on the calculated value of $\theta_W$.
\item Expansion of $\mscr{L}_b^{(\mrm{fg})}$ and substitution according to \Erefr{eq:subA}{eq:subZ} %
yields appropriate boson interaction vertices for the electroweak sector, %
supplemented by right-handed weak interactions involving $G^\bdag_\mu$.
\end{itemize}

\subsubsection{Strong interactions\label{sec:strongint}}

As noted in \sref{sec:catalogueall}, this model contains exactly the correct number of nonvanishing composite fermionic species to describe all generations of the observed leptons and quarks. Further, examination of their constituents confirms that all species carry the correct relative electromagnetic charges, and qualitatively correct electroweak interactions (with quantitative assessment necessarily awaiting a more numerically precise exploration of the electroweak sector, which begins in \cref{ch:detail}). %

In addition, when the fermionic mass mechanism is described in \cref{ch:fermion}, %
it is readily seen that the composite species identified with the up quarks will have different masses to the down quarks due to the different $A$-charges of their constituents (and likewise for their higher-generational counterparts), while the left- and right-handed members of a species have the same effective mass vertices in the low-energy limit (though, as in the Standard Model, these may be expected to run differently with energy scale due to suppression of the right-handed weak interaction).

For the quarks, mass differences between the component preons imply that although all preons are confined by the $\SU{3}_C$ interactions, the lighter component(s) (which carry an $A$-index of 3) may be expected to spend more time more distant from the common centre of mass. The colour of the $a=3$ component(s) therefore correspond to the colour of this ``halo'', and assuming shielding of the more tightly confined core, the quarks will therefore interact with a residual colour interaction corresponding to the colour (or colour deficit) of the halo, taking place by exchange of $C$-bosons $\fgfield{\bmc^\tc_\mu}$. These bosons are seen in \srefs{sec:gluonmasses}{sec:results} %
to have a hypothetical rest mass, but it is uncertain if this can ever be observed as these bosons themselves also carry a colour charge and exhibit confinement with some energy scale which may likewise be greater than $\mc{E}_0$. %

The strength of the strong interaction in CASMIR has not yet been formally determined. While the CASMIR model incorporates coloured boson interaction vertices with coupling constants of $\ILO{\bmmf_A}\sim\ILO{\la\vp\ra^{-1}}$, the characteristic length scale of strong interactions corresponds to the preon confinement scale \schapnotchap{sec:chromenv}{pfeifer2022CASM4}, and thus it is also possible that confinement is mediated by direct preon exchange without $\ILO{\bmmf_A}$ vertices, for an effective coupling constant of~1.

Further study of the colour sector is required to determine the details of the strong interaction, both at the level of preons to establish interaction strengths, and at the level of quarks, for example to see if association of quarks into colour-neutral triplets selects naturally for the observed triplets of $udd$ and $uud$ on account of gauge, and if this construction yields appropriate spins in the $\Cw{18}$ model. This is deferred to a subsequent paper.

\subsubsection{$N$~boson interactions}

The $N$~boson has essentially no interactions with the other spin-1 bosons as it commutes with all elements of $\su{3}_A$ and $\su{3}_C$. It likewise has no interactions with fermions due to gauge choices~\erefs{eq:U1gauge}{eq:GL1RNgauge} and equivalent extensions to the quark sectors when no leptons are present. An exception is made for $N$-mediated zero-range interactions when two fields which otherwise would not interact with the $N$ field are collocated, but this is of minimal physical significance.

To practical intents the $N$~boson therefore interacts only via scalar boson exchange, and via the gravitational analogue introduced in \cref{ch:gravity}.

\subsubsection{Comparison of vector boson coupling constants\label{sec:EWint_numerical}}

Given the different preon structures making up the leptons and the quarks (\srefr{sec:catalogue}{sec:quarksgluons}), and the different mechanisms in play for diagonal and off-diagonal interactions (\srefr{sec:photonint}{sec:EWint_Wintdetail}), it is useful to explicitly expand the symmetry factors of a representative selection of vector boson vertices and comfirm they yield the expected results.

\paragraph[Lepton/photon vertex]{Lepton/photon vertex:}~\\
\begin{itemize}
\item Three choices of which preon to interact with.
\item All choices are compatible with the interaction being studied.
\item Interaction strength $\bmmf_A/\sqrt{2}$.
\item Factor of $\frac{1}{3}$ from the definition of a fermion~\ereft{eq:generalfermion}{eq:collectivecompositefermions}.
\end{itemize}
Net factor: $\bmmf_A/\sqrt{2}$. Colour-agnostic: Colour distributions play no active role in the calculation and thus are averaged over.~\\

\paragraph[Quark/photon vertex]{Quark/photon vertex:}~\\
\begin{itemize}
\item Three choices of which preon to interact with.
\item All choices are compatible with the interaction being studied.
\item Mean interaction strength $\pm k\bmmf_A/(3\sqrt{2})$, $k\in\{1,2\}$ depending on species.
\item Factor of $\frac{1}{3}$ from the definition of a fermion.
\end{itemize}
Net factor: $\pm k\bmmf_A/(3\sqrt{2})$, $k\in\{1,2\}$. Colour-agnostic.~\\

Counting for the $Z$ boson is equivalent to that for the photon, and similarly yields the expected values.\\

\paragraph[Lepton/\prm{W} vertex]{Lepton/\prm{W} vertex:}
\begin{itemize}
\item Three choices of which preon to interact with.
\item All choices are compatible with the interaction being studied. 
\item Interaction strength $\bmmf_A$.
\item Factor of $\frac{1}{3}$ from the definition of a fermion.
\end{itemize}
Net factor: $\bmmf_A$. Colour-agnostic.~\\

\paragraph[Quark/W vertex]{Quark/W vertex:}
\begin{itemize}
\item Three choices (by spatial co-ordinate) of which preon to interact with.
\item Any preon may be the unique $A$-charge: Three different $A$-charge arrangements.
\item Interaction only valid if unique $A$-charge coincides with interacting preon: Factor $\frac{1}{3}$.
\item Interaction strength $\bmmf_A$.
\item Factor of $\frac{1}{3}$ from the definition of a fermion.
\end{itemize}
Net factor: $\bmmf_A$. Colour-agnostic.~\\

\paragraph[Fermion/gluon vertex (one specific gluon)]{Fermion/gluon vertex (one specific gluon):\label{sec:EWint_numerical_f/g}}
\begin{itemize}
\item Three (spatial) choices of which preon to interact with.
\item Three choices of colour on that preon.
\item For a given gluon, one-in-three chance of that preon colour being appropriate.
\item Interaction strength $\bmmf_A$.
\item Factor of $\frac{1}{3}$ from the definition of a fermion.
\end{itemize}
Net factor: $\bmmf_A$. $A$-charge-agnostic. Also agnostic with respect to the colour labels not involved in the gluon interaction.

\section{Conclusion}

This chapter has shown that under an appropriate choice of gauge, a model comprising free real scalar fields on $\Cw{18}$ exhibits an emergent particle spectrum analogous to that of the Standard Model, along with a qualitatively appropriate Lagrangian. The particles of the Standard Model are supplemented by three additional bosons, denoted $G_\mu$, $G^\dagger_\mu$, and $N_\mu$.

Of these, interactions involving $G_\mu$ and $G^\dagger_\mu$ resemble a right-handed weak interaction but are suppressed by a factor of $\ILO{10^{7}}$ %
(see \sref{sec:interactions} above), %
and under appropriate circumstances may be eliminated entirely (\sref{sec:Rwnf}). The $N$ boson is more enigmatic but also obscure, behaving as a ninth gluon with no net colour or charge. Nevertheless, it has subtle effects detectable in the higher-order particle mass calculations of \cref{ch:detail}. %
As an additional neutral current, it is reassuring to note that it is not flavour-changing.

Although this chapter has made substantial progress in demonstrating the existence of analogies between the $\Cw{18}$ model and the Standard Model, ultimately it is in the calculation of observable quantities that the quality and utility of this analogy will be assessed. For this reason, subsequent chapters demonstrate the mass mechanisms of the boson and fermion sectors, followed by the calculation of a selection of observable quantities to precisions comparable with experiment, providing a direct check on the consistency of this model with the Standard Model.

\appendix

\section{Lie group factorisation and gauge bosons\label{apdx:gpfac}}

Given a Lie group $G$, let $d_G$ denote the dimension of $G$, let $\mfk{g}$ denote the associated Lie algebra, and let $\{\mfk{g}_i\}$ represent a basis of that Lie algebra. Let the tangent space to the origin of $G$ be denoted $TG$.

Let $A$, $B$, and $C$ be Lie groups satisfying
\begin{equation}
A\cong B\otimes C,\label{eq:A=BC}
\end{equation}
from which follows
\begin{equation}
TA\cong T(B\otimes C)\cong TB\otimes TC.
\end{equation}
Any infinitesimal element of $A$ may be written in the form
\begin{equation}
e^{\rmi a^i\mfk{a}_i}
\end{equation}
for an appropriate choice of constants $a^i$, and maps to an element of $TA$. Any element of $TA$ maps to an element of $TB\otimes TC$, and therefore may be written
\begin{equation}
e^{\rmi b^i\mfk{b}_i}e^{\rmi c^j\mfk{c}_j}
\end{equation}
for sets of constants $\{b^i,c^j\}$. It therefore follows that $\{\mfk{b}_i,\mfk{c}_j\}$ generate a basis of $TA$.

Let $R_G$ denote a matrix representation of group $G$ with dimension $d_G$, and let $R_\mfk{g}$ denote the associated matrix representation of Lie algebra $\mfk{g}$, with basis $\{\mfk{g}^R_i\}$. Given a representation $R_A$ acting on a state vector $v$ with coupling coefficient $f$, any element of $R_A$ which may be written in exponential form then acts on $v$ as
\begin{equation}
\begin{split}
e^{\rmi f a^i\mfk{a}^R_i} v &= \left(e^{\rmi  a^i\mfk{a}^R_i}\right)^f v\\
&=\left(e^{\rmi b^i\mfk{b}^R_i}e^{\rmi c^j\mfk{c}^R_j}\right)^f v\\
&=e^{\rmi fb^i\mfk{b}^R_i}e^{\rmi fc^j\mfk{c}^R_j} v
\end{split}
\end{equation}
for some $a^i$, $b^i$, and $c^j$, where use has been made that $b^i\mfk{b}^R_i$ commutes with $c^j\mfk{c}^R_j$ for all $\{b^i,c^j\}$.
Promoting $A$ from a global to a local symmetry entails constructing an $A$-valued tangent bundle over some space--time manifold on which $v$ is a vector field, and granting $x$-dependence to $a^i$, $b^i$, and $c^j$. This operation is typically performed in the context of a model in which observable quantities correspond to expectation values with respect to field $v$.
To determine the gauge boson fields associated with this local symmetry, now evaluate
\begin{equation}
v^\dagger \left(e^{\rmi fa^i\mfk{a}^R_i}\right)^{-1}~\partial_\mu \left(e^{\rmi fa^i\mfk{a}^R_i} v\right)
\end{equation}
and
\begin{equation}
v^\dagger \left(e^{\rmi fc^j\mfk{c}^R_j}\right)^{-1}\left(e^{\rmi fb^i\mfk{b}^R_i}\right)^{-1}~\partial_\mu \left(e^{\rmi fb^i\mfk{b}^R_i}e^{\rmi fc^j\mfk{c}^R_j} v\right)
\end{equation}
to yield
\begin{equation}
v^\dagger D_\mu v\quad | \quad D_\mu = \partial_\mu-\rmi f\partial_\mu(a^i) \mfk{a}^R_i\label{eq:abosons}
\end{equation}
and
\begin{equation}
v^\dagger D_\mu v\quad | \quad D_\mu = \partial_\mu-\rmi f\partial_\mu(b^i) \mfk{b}^R_i-\rmi f\partial_\mu(c^i) \mfk{c}^R_i\label{eq:bcbosons}
\end{equation}
respectively. The emergent boson fields appear in \Erefs{eq:abosons}{eq:bcbosons}, %
\begin{equation}
g^i_\mu(x)=\partial_\mu[g^i(x)]\quad | \quad g\in\{a,b,c\}.
\end{equation} 
It follows that when constructing gauge bosons from a local symmetry group which admits a product structure~\eref{eq:A=BC}, there exists a freedom to choose either a total of $d_A=d_B\times d_C$ gauge bosons associated with the $d_A$-dimensional representation of Lie algebra $\mfk{a}$ \eref{eq:abosons}, or a total of $d_B+d_C$ gauge bosons associated with the $d_B$-dimensional representation of $\mfk{b}$ and the $d_C$-dimensional representation of $\mfk{c}$ \eref{eq:bcbosons}. With limited exceptions, 
\begin{equation}
d_B+d_C<d_B\times d_C. 
\end{equation}
While both choices describe the same local symmetry, the latter choice yields the more concise model.

\appendixend

\notchap{
\section*{Acknowledgements}
This research was supported in part by the Perimeter Institute for Theoretical Physics.
Research at the Perimeter Institute is supported by the Government of Canada through Industry Canada and by the Province of Ontario through the Ministry of Research and Innovation.
The author thanks the Ontario Ministry of Research and Innovation Early Researcher Awards (ER09-06-073) for financial support.
This project was supported in part through the Macquarie University Research Fellowship scheme.
This research was supported in part by the ARC Centre of Excellence in Engineered Quantum Systems (EQuS), Project No.~CE110001013.
}

%% file: arXiv3.bbl
\begin{thebibliography}{27}%
\makeatletter
\providecommand \@ifxundefined [1]{%
 \@ifx{#1\undefined}
}%
\providecommand \@ifnum [1]{%
 \ifnum #1\expandafter \@firstoftwo
 \else \expandafter \@secondoftwo
 \fi
}%
\providecommand \@ifx [1]{%
 \ifx #1\expandafter \@firstoftwo
 \else \expandafter \@secondoftwo
 \fi
}%
\providecommand \natexlab [1]{#1}%
\providecommand \enquote  [1]{``#1''}%
\providecommand \bibnamefont  [1]{#1}%
\providecommand \bibfnamefont [1]{#1}%
\providecommand \citenamefont [1]{#1}%
\providecommand \href@noop [0]{\@secondoftwo}%
\providecommand \href [0]{\begingroup \@sanitize@url \@href}%
\providecommand \@href[1]{\@@startlink{#1}\@@href}%
\providecommand \@@href[1]{\endgroup#1\@@endlink}%
\providecommand \@sanitize@url [0]{\catcode `\\12\catcode `\$12\catcode
  `\&12\catcode `\#12\catcode `\^12\catcode `\_12\catcode `\%12\relax}%
\providecommand \@@startlink[1]{}%
\providecommand \@@endlink[0]{}%
\providecommand \url  [0]{\begingroup\@sanitize@url \@url }%
\providecommand \@url [1]{\endgroup\@href {#1}{\urlprefix }}%
\providecommand \urlprefix  [0]{URL }%
\providecommand \Eprint [0]{\href }%
\providecommand \doibase [0]{https://doi.org/}%
\providecommand \selectlanguage [0]{\@gobble}%
\providecommand \bibinfo  [0]{\@secondoftwo}%
\providecommand \bibfield  [0]{\@secondoftwo}%
\providecommand \translation [1]{[#1]}%
\providecommand \BibitemOpen [0]{}%
\providecommand \bibitemStop [0]{}%
\providecommand \bibitemNoStop [0]{.\EOS\space}%
\providecommand \EOS [0]{\spacefactor3000\relax}%
\providecommand \BibitemShut  [1]{\csname bibitem#1\endcsname}%
\let\auto@bib@innerbib\@empty
\bibitem [{\citenamefont {Maynard}(2001)}]{maynard2001}%
  \BibitemOpen
  \bibfield  {author} {\bibinfo {author} {\bibfnamefont {J.~D.}\ \bibnamefont
  {Maynard}},\ }\href {https://doi.org/10.1103/RevModPhys.73.401} {\bibfield
  {journal} {\bibinfo  {journal} {Rev. Mod. Phys.}\ }\textbf {\bibinfo {volume}
  {73}},\ \bibinfo {pages} {401} (\bibinfo {year} {2001})}\BibitemShut
  {NoStop}%
\bibitem [{\citenamefont {Dragoman}\ and\ \citenamefont
  {Dragoman}(2004)}]{dragoman2004}%
  \BibitemOpen
  \bibfield  {author} {\bibinfo {author} {\bibfnamefont {D.}~\bibnamefont
  {Dragoman}}\ and\ \bibinfo {author} {\bibfnamefont {M.}~\bibnamefont
  {Dragoman}},\ }\href {https://www.springer.com/gp/book/9783540201472} {\emph
  {\bibinfo {title} {Quantum--Classical Analogies}}},\ The Frontiers
  Collection\ (\bibinfo  {publisher} {Springer-Verlag},\ \bibinfo {address}
  {Berlin Heidelberg},\ \bibinfo {year} {2004})\BibitemShut {NoStop}%
\bibitem [{\citenamefont {Lewenstein}\ \emph {et~al.}(2007)\citenamefont
  {Lewenstein}, \citenamefont {Sanpera}, \citenamefont {Ahufinger},
  \citenamefont {Damski}, \citenamefont {Sen(De)},\ and\ \citenamefont
  {Sen}}]{lewenstein2007}%
  \BibitemOpen
  \bibfield  {author} {\bibinfo {author} {\bibfnamefont {M.}~\bibnamefont
  {Lewenstein}}, \bibinfo {author} {\bibfnamefont {A.}~\bibnamefont {Sanpera}},
  \bibinfo {author} {\bibfnamefont {V.}~\bibnamefont {Ahufinger}}, \bibinfo
  {author} {\bibfnamefont {B.}~\bibnamefont {Damski}}, \bibinfo {author}
  {\bibfnamefont {A.}~\bibnamefont {Sen(De)}},\ and\ \bibinfo {author}
  {\bibfnamefont {U.}~\bibnamefont {Sen}},\ }\\\href
  {https://doi.org/10.1080/00018730701223200} {\bibfield  {journal} {\bibinfo
  {journal} {Adv. Phys.}\ }\textbf {\bibinfo {volume} {56}},\ \bibinfo {pages}
  {243} (\bibinfo {year} {2007})}\BibitemShut {NoStop}%
\bibitem [{\citenamefont {Onsager}(1944)}]{onsager1944}%
  \BibitemOpen
  \bibfield  {author} {\bibinfo {author} {\bibfnamefont {L.}~\bibnamefont
  {Onsager}},\ }\href {https://doi.org/10.1103/PhysRev.65.117} {\bibfield
  {journal} {\bibinfo  {journal} {Phys. Rev.}\ }\textbf {\bibinfo {volume}
  {65}},\ \bibinfo {pages} {117} (\bibinfo {year} {1944})}\BibitemShut
  {NoStop}%
\bibitem [{\citenamefont {Suzuki}(1976)}]{suzuki1976}%
  \BibitemOpen
  \bibfield  {author} {\bibinfo {author} {\bibfnamefont {M.}~\bibnamefont
  {Suzuki}},\ }\href {https://doi.org/10.1143/PTP.56.1454} {\bibfield
  {journal} {\bibinfo  {journal} {Prog. Theor. Phys.}\ }\textbf {\bibinfo
  {volume} {56}},\ \bibinfo {pages} {1454} (\bibinfo {year}
  {1976})}\BibitemShut {NoStop}%
\bibitem [{\citenamefont {Visser}\ \emph {et~al.}(2002)\citenamefont {Visser},
  \citenamefont {Barcel{\'o}},\ and\ \citenamefont {Liberati}}]{visser2002}%
  \BibitemOpen
  \bibfield  {author} {\bibinfo {author} {\bibfnamefont {M.}~\bibnamefont
  {Visser}}, \bibinfo {author} {\bibfnamefont {C.}~\bibnamefont
  {Barcel{\'o}}},\ and\ \bibinfo {author} {\bibfnamefont {S.}~\bibnamefont
  {Liberati}},\ }\href {https://doi.org/10.1023/A:1020180409214} {\bibfield
  {journal} {\bibinfo  {journal} {Gen. Rel. and Grav.}\ }\textbf {\bibinfo
  {volume} {34}},\ \bibinfo {pages} {1719} (\bibinfo {year}
  {2002})}\BibitemShut {NoStop}%
\bibitem [{\citenamefont {Liberati}\ \emph {et~al.}(2009)\citenamefont
  {Liberati}, \citenamefont {Girelli},\ and\ \citenamefont
  {Sindoni}}]{liberati2009}%
  \BibitemOpen
  \bibfield  {author} {\bibinfo {author} {\bibfnamefont {S.}~\bibnamefont
  {Liberati}}, \bibinfo {author} {\bibfnamefont {F.}~\bibnamefont {Girelli}},\
  and\ \bibinfo {author} {\bibfnamefont {L.}~\bibnamefont {Sindoni}},\
  }\href@noop {} {\bibinfo {title} {Analogue models for emergent gravity}},\
  \bibinfo {howpublished}
  {\href{https://arxiv.org/abs/0909.3834v1}{arXiv:0909.3834v1 [gr-qc]}}
  (\bibinfo {year} {2009})\BibitemShut {NoStop}%
\bibitem [{\citenamefont {Barcel{\'{o}}}\ \emph {et~al.}(2011)\citenamefont
  {Barcel{\'{o}}}, \citenamefont {Liberati},\ and\ \citenamefont
  {Visser}}]{barcelo2011}%
  \BibitemOpen
  \bibfield  {author} {\bibinfo {author} {\bibfnamefont {C.}~\bibnamefont
  {Barcel{\'{o}}}}, \bibinfo {author} {\bibfnamefont {S.}~\bibnamefont
  {Liberati}},\ and\ \bibinfo {author} {\bibfnamefont {M.}~\bibnamefont
  {Visser}},\ }\href {https://doi.org/10.12942/lrr-2011-3} {\bibfield
  {journal} {\bibinfo  {journal} {Living Reviews in Relativity}\ }\textbf
  {\bibinfo {volume} {14}},\ \bibinfo {pages} {3} (\bibinfo {year}
  {2011})}\BibitemShut {NoStop}%
\bibitem [{\citenamefont {Unruh}(1981)}]{unruh1981}%
  \BibitemOpen
  \bibfield  {author} {\bibinfo {author} {\bibfnamefont {W.~G.}\ \bibnamefont
  {Unruh}},\ }\href {https://doi.org/10.1103/PhysRevLett.46.1351} {\bibfield
  {journal} {\bibinfo  {journal} {Phys. Rev. Lett.}\ }\textbf {\bibinfo
  {volume} {46}},\ \bibinfo {pages} {1351} (\bibinfo {year}
  {1981})}\BibitemShut {NoStop}%
\bibitem [{\citenamefont {Garay}\ \emph {et~al.}(2000)\citenamefont {Garay},
  \citenamefont {Anglin}, \citenamefont {Cirac},\ and\ \citenamefont
  {Zoller}}]{garay2000}%
  \BibitemOpen
  \bibfield  {author} {\bibinfo {author} {\bibfnamefont {L.~J.}\ \bibnamefont
  {Garay}}, \bibinfo {author} {\bibfnamefont {J.~R.}\ \bibnamefont {Anglin}},
  \bibinfo {author} {\bibfnamefont {J.~I.}\ \bibnamefont {Cirac}},\ and\
  \bibinfo {author} {\bibfnamefont {P.}~\bibnamefont {Zoller}},\ }\href
  {https://doi.org/10.1103/PhysRevLett.85.4643} {\bibfield  {journal} {\bibinfo
   {journal} {Phys. Rev. Lett.}\ }\textbf {\bibinfo {volume} {85}},\ \bibinfo
  {pages} {4643} (\bibinfo {year} {2000})}\BibitemShut {NoStop}%
\bibitem [{\citenamefont {Garay}\ \emph {et~al.}(2001)\citenamefont {Garay},
  \citenamefont {Anglin}, \citenamefont {Cirac},\ and\ \citenamefont
  {Zoller}}]{garay2001}%
  \BibitemOpen
  \bibfield  {author} {\bibinfo {author} {\bibfnamefont {L.~J.}\ \bibnamefont
  {Garay}}, \bibinfo {author} {\bibfnamefont {J.~R.}\ \bibnamefont {Anglin}},
  \bibinfo {author} {\bibfnamefont {J.~I.}\ \bibnamefont {Cirac}},\ and\
  \bibinfo {author} {\bibfnamefont {P.}~\bibnamefont {Zoller}},\ }\href
  {https://doi.org/10.1103/PhysRevA.63.023611} {\bibfield  {journal} {\bibinfo
  {journal} {Phys. Rev. A}\ }\textbf {\bibinfo {volume} {63}},\ \bibinfo
  {pages} {023611} (\bibinfo {year} {2001})}\BibitemShut {NoStop}%
\bibitem [{\citenamefont {Lahav}\ \emph {et~al.}(2010)\citenamefont {Lahav},
  \citenamefont {Itah}, \citenamefont {Blumkin}, \citenamefont {Gordon},
  \citenamefont {Rinott}, \citenamefont {Zayats},\ and\ \citenamefont
  {Steinhauer}}]{lahav2010}%
  \BibitemOpen
  \bibfield  {author} {\bibinfo {author} {\bibfnamefont {O.}~\bibnamefont
  {Lahav}}, \bibinfo {author} {\bibfnamefont {A.}~\bibnamefont {Itah}},
  \bibinfo {author} {\bibfnamefont {A.}~\bibnamefont {Blumkin}}, \bibinfo
  {author} {\bibfnamefont {C.}~\bibnamefont {Gordon}}, \bibinfo {author}
  {\bibfnamefont {S.}~\bibnamefont {Rinott}}, \bibinfo {author} {\bibfnamefont
  {A.}~\bibnamefont {Zayats}},\ and\ \bibinfo {author} {\bibfnamefont
  {J.}~\bibnamefont {Steinhauer}},\ }\\\href
  {http://dx.doi.org/10.1103/PhysRevLett.105.240401} {\bibfield  {journal}
  {\bibinfo  {journal} {Phys. Rev. Lett.}\ }\textbf {\bibinfo {volume} {105}},\
  \bibinfo {pages} {240401} (\bibinfo {year} {2010})}\BibitemShut {NoStop}%
\bibitem [{\citenamefont {Gordon}(1923)}]{gordon1923}%
  \BibitemOpen
  \bibfield  {author} {\bibinfo {author} {\bibfnamefont {W.}~\bibnamefont
  {Gordon}},\ }\href {https://doi.org/10.1002/andp.19233772202} {\bibfield
  {journal} {\bibinfo  {journal} {Ann. Phys.}\ }\textbf {\bibinfo {volume}
  {377}},\ \bibinfo {pages} {421} (\bibinfo {year} {1923})}\BibitemShut
  {NoStop}%
\bibitem [{\citenamefont {Leonhardt}\ and\ \citenamefont
  {Piwnicki}(1999)}]{leonhardt1999}%
  \BibitemOpen
  \bibfield  {author} {\bibinfo {author} {\bibfnamefont {U.}~\bibnamefont
  {Leonhardt}}\ and\ \bibinfo {author} {\bibfnamefont {P.}~\bibnamefont
  {Piwnicki}},\ }\href {https://doi.org/10.1103/PhysRevA.60.4301} {\bibfield
  {journal} {\bibinfo  {journal} {Phys. Rev. A}\ }\textbf {\bibinfo {volume}
  {60}},\ \bibinfo {pages} {4301} (\bibinfo {year} {1999})}\BibitemShut
  {NoStop}%
\bibitem [{\citenamefont {Leonhardt}\ and\ \citenamefont
  {Piwnicki}(2000)}]{leonhardt2000}%
  \BibitemOpen
  \bibfield  {author} {\bibinfo {author} {\bibfnamefont {U.}~\bibnamefont
  {Leonhardt}}\ and\ \bibinfo {author} {\bibfnamefont {P.}~\bibnamefont
  {Piwnicki}},\ }\href {https://doi.org/10.1103/PhysRevLett.84.822} {\bibfield
  {journal} {\bibinfo  {journal} {Phys. Rev. Lett.}\ }\textbf {\bibinfo
  {volume} {84}},\ \bibinfo {pages} {822} (\bibinfo {year} {2000})}\BibitemShut
  {NoStop}%
\bibitem [{\citenamefont {Jacobson}\ and\ \citenamefont
  {Volovik}(1998)}]{jacobson1998}%
  \BibitemOpen
  \bibfield  {author} {\bibinfo {author} {\bibfnamefont {T.~A.}\ \bibnamefont
  {Jacobson}}\ and\ \bibinfo {author} {\bibfnamefont {G.~E.}\ \bibnamefont
  {Volovik}},\ }\href {https://doi.org/10.1103/PhysRevD.58.064021} {\bibfield
  {journal} {\bibinfo  {journal} {Phys. Rev. D}\ }\textbf {\bibinfo {volume}
  {58}},\ \bibinfo {pages} {064021} (\bibinfo {year} {1998})}\BibitemShut
  {NoStop}%
\bibitem [{\citenamefont {Reznik}(2000)}]{reznik2000}%
  \BibitemOpen
  \bibfield  {author} {\bibinfo {author} {\bibfnamefont {B.}~\bibnamefont
  {Reznik}},\ }\href {https://doi.org/10.1103/PhysRevD.62.044044} {\bibfield
  {journal} {\bibinfo  {journal} {Phys. Rev. D}\ }\textbf {\bibinfo {volume}
  {62}},\ \bibinfo {pages} {044044} (\bibinfo {year} {2000})}\BibitemShut
  {NoStop}%
\bibitem [{\citenamefont {Sch\"utzhold}\ and\ \citenamefont
  {Unruh}(2005)}]{schutzhold2005}%
  \BibitemOpen
  \bibfield  {author} {\bibinfo {author} {\bibfnamefont {R.}~\bibnamefont
  {Sch\"utzhold}}\ and\ \bibinfo {author} {\bibfnamefont {W.~G.}\ \bibnamefont
  {Unruh}},\ }\href {https://doi.org/10.1103/PhysRevLett.95.031301} {\bibfield
  {journal} {\bibinfo  {journal} {Phys. Rev. Lett.}\ }\textbf {\bibinfo
  {volume} {95}},\ \bibinfo {pages} {031301} (\bibinfo {year}
  {2005})}\BibitemShut {NoStop}%
\bibitem [{\citenamefont {Sch\"utzhold}\ and\ \citenamefont
  {Unruh}(2002)}]{schutzhold2002}%
  \BibitemOpen
  \bibfield  {author} {\bibinfo {author} {\bibfnamefont {R.}~\bibnamefont
  {Sch\"utzhold}}\ and\ \bibinfo {author} {\bibfnamefont {W.~G.}\ \bibnamefont
  {Unruh}},\ }\href {https://doi.org/10.1103/PhysRevD.66.044019} {\bibfield
  {journal} {\bibinfo  {journal} {Phys. Rev. D}\ }\textbf {\bibinfo {volume}
  {66}},\ \bibinfo {pages} {044019} (\bibinfo {year} {2002})}\BibitemShut
  {NoStop}%
\bibitem [{\citenamefont {Aguillard}\ \emph {et~al.}(2023)\citenamefont
  {Aguillard} \emph {et~al.}}]{aguillard2023}%
  \BibitemOpen
  \bibfield  {author} {\bibinfo {author} {\bibfnamefont {D.~P.}\ \bibnamefont
  {Aguillard}} \emph {et~al.} (\bibinfo {collaboration} {The Muon $g-2$
  Collaboration}),\ }\href@noop {} {\bibinfo {title} {Measurement of the
  positive muon anomalous magnetic moment to 0.20 ppm}},\ \bibinfo
  {howpublished} {\href{https://arxiv.org/abs/2308.06230v1}{arXiv:2308.06230v1
  [hep-ex]}} (\bibinfo {year} {2023})\BibitemShut {NoStop}%
\bibitem [{\citenamefont {Aoyama}\ \emph {et~al.}(2020)\citenamefont {Aoyama}
  \emph {et~al.}}]{aoyama2020}%
  \BibitemOpen
  \bibfield  {author} {\bibinfo {author} {\bibfnamefont {T.}~\bibnamefont
  {Aoyama}} \emph {et~al.},\ }\href
  {https://doi.org/10.1016/j.physrep.2020.07.006} {\bibfield  {journal}
  {\bibinfo  {journal} {Physics Reports}\ }\textbf {\bibinfo {volume} {887}},\
  \bibinfo {pages} {1} (\bibinfo {year} {2020})}\BibitemShut {NoStop}%
\bibitem [{\citenamefont {Borsanyi}\ \emph {et~al.}(2021)\citenamefont
  {Borsanyi} \emph {et~al.}}]{borsanyi2021}%
  \BibitemOpen
  \bibfield  {author} {\bibinfo {author} {\bibfnamefont {S.}~\bibnamefont
  {Borsanyi}} \emph {et~al.},\ }\href
  {https://doi.org/10.1038/s41586-021-03418-1} {\bibfield  {journal} {\bibinfo
  {journal} {Nature}\ }\textbf {\bibinfo {volume} {593}},\ \bibinfo {pages}
  {51} (\bibinfo {year} {2021})}\BibitemShut {NoStop}%
\bibitem [{\citenamefont {Colangelo}\ \emph {et~al.}(2022)\citenamefont
  {Colangelo} \emph {et~al.}}]{colangelo2022}%
  \BibitemOpen
  \bibfield  {author} {\bibinfo {author} {\bibfnamefont {G.}~\bibnamefont
  {Colangelo}} \emph {et~al.},\ }\href@noop {} {\bibinfo {title} {Prospects for
  precise predictions of $a_\mu$ in the {Standard} {Model}}},\ \bibinfo
  {howpublished} {\href{https://arxiv.org/abs/2203.15810v1}{arXiv:2203.15810v1
  [hep-ph]}} (\bibinfo {year} {2022})\BibitemShut {NoStop}%
\bibitem [{\citenamefont {Ignatov}\ \emph {et~al.}(2023)\citenamefont {Ignatov}
  \emph {et~al.}}]{ignatov2023}%
  \BibitemOpen
  \bibfield  {author} {\bibinfo {author} {\bibfnamefont {F.~V.}\ \bibnamefont
  {Ignatov}} \emph {et~al.} (\bibinfo {collaboration} {{CMD-3}
  Collaboration}),\ }\href@noop {} {\bibinfo {title} {Measurement of the
  $e^+e^-\rightarrow\pi^+\pi^-$ cross-section from threshold to
  {$1.2~\mrm{GeV}$} with the {CMD-3} detector}},\ \bibinfo {howpublished}
  {\href{https://arxiv.org/abs/2302.08834v2}{arXiv:2302.08834v2 [hep-ex]}}
  (\bibinfo {year} {2023})\BibitemShut {NoStop}%
\bibitem [{\citenamefont {Ryder}(1996)}]{ryder1996}%
  \BibitemOpen
  \bibfield  {author} {\bibinfo {author} {\bibfnamefont {L.~H.}\ \bibnamefont
  {Ryder}},\ }\href@noop {} {\emph {\bibinfo {title} {Quantum Field Theory}}},\
  \bibinfo {edition} {2nd}\ ed.\ (\bibinfo  {publisher} {University Press},\
  \bibinfo {address} {Cambridge},\ \bibinfo {year} {1996})\BibitemShut
  {NoStop}%
\bibitem [{\citenamefont {Tiesinga}\ \emph {et~al.}(2018)\citenamefont
  {Tiesinga}, \citenamefont {Mohr}, \citenamefont {Newell},\ and\ \citenamefont
  {Taylor}}]{tiesinga2018}%
  \BibitemOpen
  \bibfield  {author} {\bibinfo {author} {\bibfnamefont {E.}~\bibnamefont
  {Tiesinga}}, \bibinfo {author} {\bibfnamefont {P.~J.}\ \bibnamefont {Mohr}},
  \bibinfo {author} {\bibfnamefont {D.~B.}\ \bibnamefont {Newell}},\ and\
  \bibinfo {author} {\bibfnamefont {B.~N.}\ \bibnamefont {Taylor}}} (\bibinfo
  {year} {2018}),\ \bibinfo {note} {({Web} {Version} 8.1). Database developed
  by J.~Baker, M.~Douma, and S.~Kotochigova. Available at
  \href{http://physics.nist.gov/constants}{http://physics.nist.gov/constants},
  National Institute of Standards and Technology, Gaithersburg, MD
  20899}\BibitemShut {NoStop}%
\bibitem [{\citenamefont {Pfeifer}(2023)}]{pfeifer2022CASM4}%
  \BibitemOpen
  \bibfield  {author} {\bibinfo {author} {\bibfnamefont {R.~N.~C.}\
  \bibnamefont {Pfeifer}},\ }
  {\bibinfo {title} {A classical analogue to the {Standard} {Model}, chapters
  4-10: {Particle} generations and masses, curved spacetimes and
  gravitation}},\ \bibinfo {howpublished} {\href {https://doi.org/10.48550/arXiv.2008.05893}{arXiv:2008.05893} with updates at
  \href{https://www.academia.edu/65931513}{https://www.academia.edu/65931513}}
  (\bibinfo {year} {2023})\BibitemShut {NoStop}%
\end{thebibliography}%
